\documentclass[longauth]{aa}

\usepackage{graphicx}
\usepackage{txfonts}
\usepackage{gensymb}
\usepackage{multicol}
\usepackage{hyperref}

\usepackage{xcolor}

\usepackage{placeins}

\begin{document} 

\title{Molecular gas and star formation in central rings across nearby galaxies}

\newcommand{\MPIA}{\label{MPIA} Max-Planck-Institut f\"{u}r Astronomie, K\"{o}nigstuhl 17, D-69117, Heidelberg, Germany}
\newcommand{\NAOJ}{\label{NAOJ} National Astronomical Observatory of Japan, 2-21-1 Osawa, Mitaka, Tokyo 181-8588, Japan}
\newcommand{\UGent}{\label{UGent} Sterrenkundig Observatorium, Universiteit Gent, Krijgslaan 281 S9, B-9000 Gent, Belgium}
\newcommand{\OAN}{\label{OAN} Observatorio Astron\'{o}mico Nacional (IGN), C/Alfonso XII, 3, E-28014 Madrid, Spain}
\newcommand{\ESO}{\label{ESO} European Southern Observatory, Karl-Schwarzschild Stra{\ss}e 2, D-85748 Garching bei M\"{u}nchen, Germany}
\newcommand{\ULyon}{\label{ULyon} Univ Lyon, Univ Lyon 1, ENS de Lyon, CNRS, Centre de Recherche Astrophysique de Lyon UMR5574,\\ F-69230 Saint-Genis-Laval, France}
\newcommand{\OSU}{\label{OSU} Department of Astronomy, The Ohio State University, 140 West 18th Avenue, Columbus, Ohio 43210, USA}
\newcommand{\UBonn}{\label{UBonn} Argelander-Institut f\"ur Astronomie, Universit\"at Bonn, Auf dem H\"ugel 71, 53121 Bonn, Germany}
\newcommand{\Alberta}{\label{Alberta} Department of Physics, University of Alberta, Edmonton, AB T6G 2E1, Canada}
\newcommand{\ITA}{\label{ITA} Universit\"{a}t Heidelberg, Zentrum f\"{u}r Astronomie, Institut f\"{u}r Theoretische Astrophysik, Albert-Ueberle-Str 2, D-69120 Heidelberg, Germany}
\newcommand{\COOL}{\label{COOL}{Cosmic Origins Of Life (COOL) Research DAO, \href{https://coolresearch.io}{https://coolresearch.io}}}
\newcommand{\UWyoming}{\label{UWyoming} Department of Physics and Astronomy, University of Wyoming, Laramie, WY 82071, USA}
\newcommand{\ANU}{\label{ANU} Research School of Astronomy and Astrophysics, Australian National University, Canberra, ACT 2611, Australia}
\newcommand{\IWR}{\label{IWR} Universit\"{a}t Heidelberg, Interdisziplin\"{a}res Zentrum f\"{u}r Wissenschaftliches Rechnen, Im Neuenheimer Feld 205, D-69120 Heidelberg, Germany}
\newcommand{\CfA}{\label{CfA} Harvard-Smithsonian Center for Astrophysics, 60 Garden Street, Cambridge, MA 02138, USA}
\newcommand{\Radcliffe}{\label{Radcliffe}{Elizabeth S.\ and Richard M.\ Cashin Fellow at the Radcliffe Institute for Advanced Studies at Harvard University, 10 Garden Street, Cambridge, MA 02138, USA}}
\newcommand{\STScI}{\label{STScI} Space Telescope Science Institute, 3700 San Martin Drive, Baltimore, MD 21218, USA}
\newcommand{\TKU}{\label{TKU} Department of Physics, Tamkang University, No.151, Yingzhuan Rd., Tamsui Dist., New Taipei City 251301, Taiwan}
\newcommand{\UCM}{\label{UCM} Departamento de F\'{\i}sica de la Tierra y Astrof\'{\i}sica, Universidad Complutense de Madrid, E-28040, Spain}
\newcommand{\Princeton}{\label{Princeton} Department of Astrophysical Sciences, Princeton University, 4 Ivy Ln., Princeton, NJ 08544 USA}
\newcommand{\Maryland}{\label{Maryland} Department of Astronomy, University of Maryland, 4296 Stadium Drive, College Park, MD 20742, USA}
\newcommand{\Oxford}{\label{Oxford} Sub-department of Astrophysics, Department of Physics, University of Oxford, Keble Road, Oxford OX1 3RH, UK}
\newcommand{\insubria}{ \label{insubria} Como Lake centre for AstroPhysics (CLAP), DiSAT, Universit{\`a} dell’Insubria, via Valleggio 11, 22100 Como, Italy}

\author{
Damian~R.~Gleis\inst{\ref{MPIA}} 
\and Sophia~K.~Stuber\inst{\ref{MPIA}, \ref{NAOJ}} 
\and Eva~Schinnerer\inst{\ref{MPIA}}
\and Justus~Neumann\inst{\ref{MPIA}} %
\and Sharon~E.~Meidt\inst{\ref{UGent}}        
\and Miguel~Querejeta\inst{\ref{OAN}} 
\and Eric~Emsellem\inst{\ref{ESO},\ref{ULyon}}
\and Adam~K.~Leroy\inst{\ref{OSU}}            
\and Ashley~T.~Barnes\inst{\ref{ESO}}
\and Frank~Bigiel\inst{\ref{UBonn}} 
\and Charlie~Burton\inst{\ref{Alberta}} 
\and Mélanie~Chevance\inst{\ref{ITA},\ref{COOL}} 
\and Daniel~A.~Dale\inst{\ref{UWyoming}} 
\and Kathryn~Grasha\inst{\ref{ANU}}
\and Ralf~S.~Klessen\inst{\ref{ITA},\ref{IWR},\ref{CfA},\ref{Radcliffe}} 
\and Rebecca~C.~Levy\inst{\ref{STScI}} 
\and Lukas~Neumann\inst{\ref{ESO}} 
\and Hsi-An~Pan\inst{\ref{TKU}} 
\and Marina~Ruiz-Garc\'ia\inst{\ref{OAN}}\inst{\ref{UCM}} 
\and Mattia~C.~Sormani\inst{\ref{insubria}}
\and Jiayi~Sun\inst{\ref{Princeton}} 
\and Yu-Hsuan~Teng\inst{\ref{Maryland}} 
\and Thomas~G.~Williams\inst{\ref{Oxford}}
}
\institute{\MPIA{}
\and \NAOJ{}
\and \UGent{}
\and \OAN{}
\and \ESO{}
\and \ULyon{}
\and \OSU{}
\and \UBonn{}
\and \Alberta{}
\and \ITA{}
\and \COOL{}
\and \UWyoming{}
\and \ANU{}
\and \IWR{}
\and \CfA{}
\and \Radcliffe{}
\and \STScI{}
\and \TKU{}
\and \UCM{}
\and \insubria{}
\and \Princeton{}
\and \Maryland{}
\and \Oxford{}
}

\date{Received ..... / accepted .....}

    \abstract
    {Nearby galaxies exhibit a variety of structures, including so-called central or (circum-)nuclear rings that are similar to the Milky Way (MW) Central Molecular Zone (CMZ). These rings are common in barred galaxies and can be gas-rich and highly star-forming.}
    {We aim to study the molecular gas content and star formation rate of central rings within nearby galaxies and link them to global galaxy properties -- especially  the bar morphology.}
    {We utilize $1\,$\arcsec (${\lesssim} 100\,$pc) resolution CO(2-1) observations from the PHANGS-ALMA survey, visually identify 20 central rings and determine their properties. For $14$ of these rings, MUSE observations tracing star formation rate (SFR) surface density are available. We derive the rings' geometry, integrated molecular gas masses, SFRs, depletion times, and compare them to host galaxy and bar properties from the literature.}
    {Molecular gas is an effective tracer for central rings: Previous studies used ionized gas and dust tracers to identify central rings in galaxies of similar morphological types as the PHANGS galaxies (numerical Hubble type  $T{\sim}-3$ to $T{\sim}9$). Compared with those, molecular gas yields similar  fractions of galaxies hosting central rings and similar radii distributions. The gaseous central rings have typical radii of ${\sim} 400_{-150}^{+250}\,$pc, molecular gas masses of $\log(M_\text{mol}/M_\odot){\sim}8.1_{-0.23}^{+0.17}$, and SFRs of ${\sim} 0.21_{-0.16}^{+0.15}\,M_\odot/\text{yr}$, thus contributing $5.6_{-2.1}^{+4.5}\,\%$ and $13_{-5}^{+10}\,\%$ to their host galaxies' molecular gas mass and SFR, respectively. While the MW CMZ sits at the lower end of the radius, molecular gas mass, and SFR distribution, it matches well in terms of ring molecular gas mass and SFR fraction, and depletion time. Longer bars contain more massive molecular central rings, but there is no correlation between the classical bar strength parameters ($Q_\text{b}$, $\varepsilon_\text{bar}$, $A_2^\text{max}$) and the ring's molecular gas content.}
    {Although absolute central ring properties (ring radius, molecular gas mass, SFR) likely depend on host galaxy properties, the similarities between the MW CMZ and PHANGS central rings in relative parameters (molecular gas and SFR fraction, depletion time) suggest that the processes of gas inflow and star formation are similar for central rings across nearby galaxies.}

   \keywords{galaxies: structure -- galaxies: ISM -- galaxies: star formation -- ISM: molecules -- Galaxy: center}

\maketitle

\section{Introduction}\label{sec:intro}

Galaxies in the nearby universe display a large variety of structures such as spiral arms, bars, and various kinds of rings in their stellar and gaseous distributions \citep[e.g.,][and many others]{kormendy_kennicutt_2004, BT08_galdyn, Salo_2015, stuber_gas_2023}. Here we study extra-galactic central rings\footnote{Central rings are also often referred to as  ``nuclear'' or  ``circumnuclear'' rings, but we use the term  ``central rings'' as the formation of such rings is thought to be linked to larger global features like bars rather than to nuclear activity.} which are common features, especially in the centers of barred galaxies \citep{knapen_2005, comeron_ainur_2010, stuber_gas_2023, erwin_2024_nuc_rings}. 
Central rings have typical radii of a few $100\,$pc, and so far they have been mostly identified as ring-like structures in the near-infrared, optical, or UV \citep[e.g. via H$\alpha$, Pa$\alpha$:][]{maoz_UV-rings_1996, knapen_2005, comeron_ainur_2010}. Central rings can have high surface densities of gas, young stars and star formation, they are subject to high rotational shears and might experience stellar and active galactic nucleus (AGN) feedback \citep{schinner_2024}. Thus, they serve as examples for the most extreme environments for star formation \citep[e.g.,][]{levy_2022_ngc253}. Furthermore, they may cause (gas) outflows affecting the evolution of their host galaxy in a substantial way \citep{leaman_2019_ring_outflow, veilleux_2020, nguyen_2022_ring_outflow}.

The Milky Way (MW) is thought to host a ring-like structure in its gas distribution with a radius of ${\sim} 100{-}200\,$pc, which is known as Central Molecular Zone (CMZ) \citep[e.g.,][]{molinari_2011_cmz, kruijssen_2015_cmz_model, henshaw_2016_cmz_geo, henshaw_2016_cmz_kin, battersby_2025_3D-CMZ_I, battersby_2025_3D-CMZ_II}. Our location inside the galactic disk makes it difficult to determine the CMZ's exact extent and to properly connect its several gas components (clouds) leaving its exact 3D geometry under debate \citep{kruijssen_2015_cmz_model, henshaw_2016_cmz_geo, Henshaw_2023_CMZ, walker_2025_3D-CMZ_III, lipman_2025_3D-CMZ_IV}. 
Thus, studying extra-galactic central rings is both interesting in its own right and useful to better understand the MW CMZ. Among many topics, it allows for examining star formation under extreme conditions, dynamical processes in the central regions of galaxies, and the connection between large-scale galaxy structure and star formation processes. Lastly, it can provide insights for the interpretation of the MW CMZ data. In this study, we determine integrated properties of central rings, such as molecular gas  masses and SFRs, and connect these with global properties of their host galaxies, such as their position on the star forming main sequence (SFMS) of galaxies  \citep{Brinchmann_2004_SFMS, Noeske_2007}, and bar properties.

Bars are thought to form due to dynamical instabilities in rotationally supported disks \citep{sellwood_bars_1993} or due to (tidal) interactions of smooth rotating disks with other galaxies or a triaxial dark matter halo \citep{noguchi_bars_1987, kormendy_kennicutt_2004, lokas_bar_2014, lokas_bars_2016}, and have typical sizes (semi-major axis) of $1{-}5\,$kpc \citep{diaz-garcia_characterization_2016}. They are mostly made up by the family of the commonly called $x_1$-orbits stabilized by the bar's self-gravity \citep{contopoulos_x1_1980}. Depending on the shape of the disk potential and the bar pattern speed, resonances like the inner Lindblad resonance (ILR) can exist \citep[e.g.,][]{BT08_galdyn, ruiz-garcia_2024_ILR_OLR} that have been linked to the presence of central rings  \citep[e.g.,][]{schwarz_1984_ILR, athanassoula_1992, combes_1988_rings, combes_1996_rings, buta_1996_rings, sormani_2024}. Molecular gas is often detected in bar lanes (or dust lanes) along the leading edge of the bar \citep[e.g.,][]{Schinnerer_2023}. These lanes are the loci of shocks and lead to gas flows towards the center (e.g.\ \citealt{athanassoula_1992, sormani_2015_bar-flow}). It has been suggested that strong bars are more efficient in bringing gas to the center (e.g., \citealt{schwarz_bars_1981, sellwood_bars_1993}). The  bar strength is often characterized via the non-axisymmetric torque parameter $Q_\text{b}$, which is the maximum normalized tangential force amplitude within the bar region  \citep[e.g.,][]{laurikainen_2002_Qb, diaz-garcia_characterization_2016}. Other measures of the bar strengths are bar ellipticity and  Fourier amplitudes of a Fourier decomposition of the images (see \citealt{diaz-garcia_characterization_2016}, and references within).

It is interesting to investigate connections between bar parameters and central ring properties, as there is still no full consensus on the formation mechanisms for these rings. Most theories on central ring formation agree that (a) non-axisymmetric structures such as bars funnel gas to the center where (b) their non-axisymmetric potential plays an important role in rearranging the matter such that central rings are formed \citep[e.g.,][]{sormani_2024}. Therefore, central rings are mostly expected within bars, although strong spiral arms or interactions with companion galaxies may induce ring formation too \citep[e.g.,][]{comeron_ainur_2010}. Central rings are traditionally thought to form at or near the ILR, typically composed by the family of $x_2$-orbits\footnote{This family of orbits lie at or within the ILR, if one is present -- not all barred galaxies show an ILR, see, e.g., \citet{athanassoula_1992}.  If two ILRs are present, the $x_2$-orbits lie between them \citep[e.g.][]{athanassoula_1992_bar_orbits, buta_1996_rings}.} in non-axisymmetric (bar) potentials (e.g. \citealt{buta_1996_rings, li_2015_ring_sim, sormani_2024}).

High-resolution radio images (${\sim} 1\arcsec \simeq 100\,$pc at a distance of $20\,$Mpc) of galaxy centers have been obtained by several studies focusing on individual galaxies or small galaxy samples \citep[e.g.,][]{garcia-burillo_NUGA_2003, callanan_2021_m83_cmz, sun_2024_ngc3351_highres}. Recently, the PHANGS-ALMA large program (Physics at High Angular resolution in Nearby GalaxieS\footnote{\url{http://phangs.org/}}; \citealt{Leroy_2021}) has provided comparable images for a large sample (90) of nearby galaxies \citep{Leroy_2021}. \citet{stuber_gas_2023} used the PHANGS-ALMA data to study the CO morphology of nearby galaxies, identifying central rings in ${\sim} 30\,\%$ of these galaxies, the majority of which are barred. With the same data set, it is possible for the first time to study these molecular central rings in detail, yielding molecular gas masses and star formation rates based on the CO ring geometry. This also provides observational constraints for theories of central ring formation and the statistics needed for studying systems similar to the MW CMZ.

The structure of this paper is as follows: In Sect.\,\ref{sec:data}, we present the data from the PHANGS survey used in this study, and define our sample of central rings. The analysis procedure is explained in detail (Sect.\,\ref{sec:analysis}) before moving on to the results in Sect.\,\ref{sec:results}. We present several properties of PHANGS central rings, compare them with the MW CMZ properties, and investigate correlations between literature bar parameters and central ring molecular gas mass. Finally, in Sect.\,\ref{sec:discuss}, we discuss the results and some caveats of this study. A conclusion can be found in Sect.\,\ref{sec:conclusion}.

\section{Data}\label{sec:data}

\begin{figure*}
\sidecaption
  \includegraphics[width=12cm]{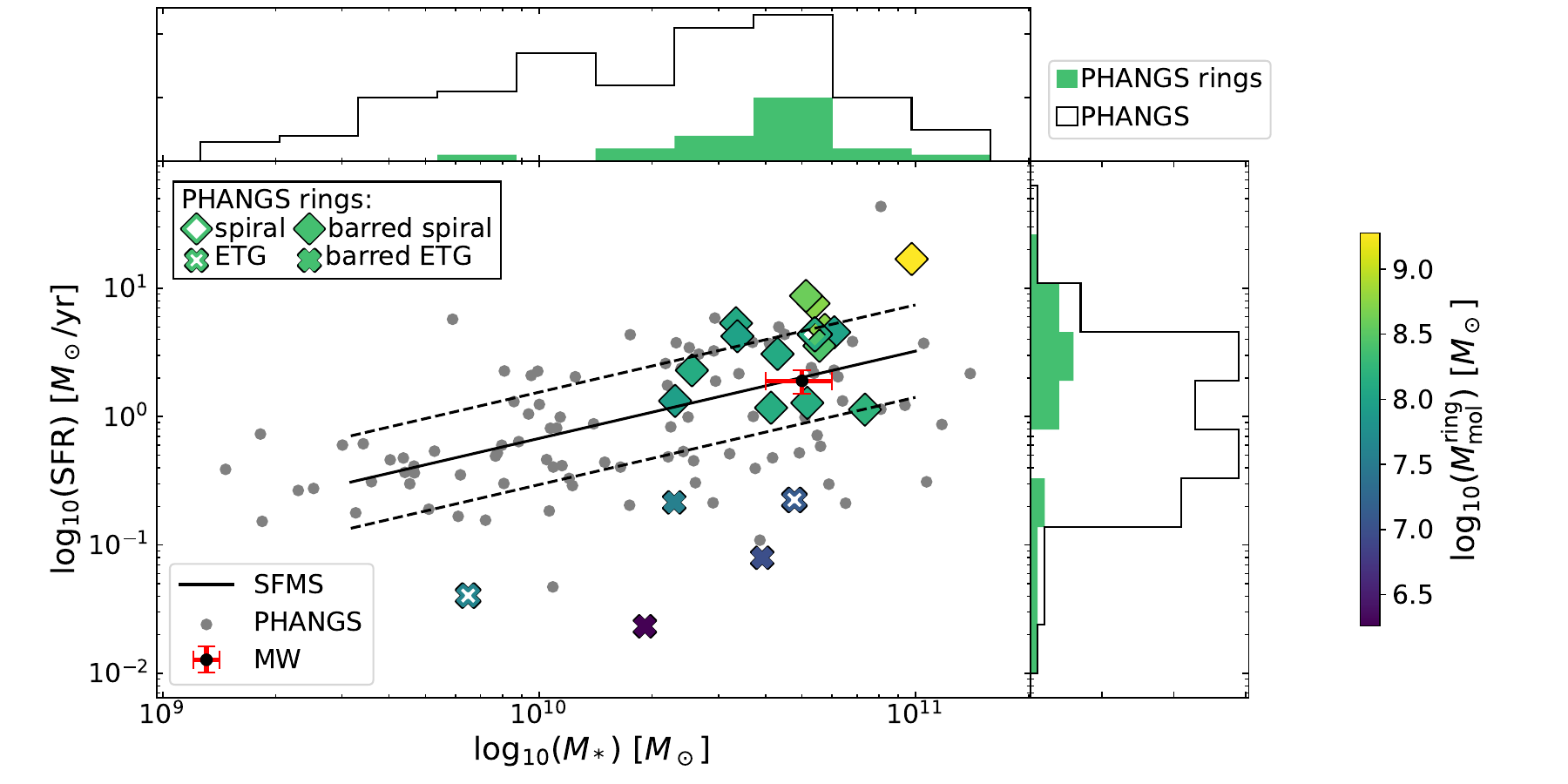}
     \caption{Global SFR as a function of global stellar mass for our sample galaxies from the PHANGS-ALMA sample. The solid and dashed black lines denote the main sequence of star forming galaxies with a scatter of 0.36\,dex, as given by \citet{Leroy_2019}. The PHANGS galaxies with central rings are color coded by the central ring gas mass as obtained in this study and the different symbols denote barred spiral (diamond) and unbarred spiral (open diamond) galaxies, barred early-type galaxies (ETGs, cross) and unbarred ETGs (open cross). The MW is plotted as a black dot with red errorbars.}
     \label{fig:rings_on_SFMS}
\end{figure*}

The PHANGS survey provides the ideal data set for studying star formation processes in a variety of structures in galaxies with different morphologies.
PHANGS investigates $90$ nearby galaxies associated with the $z=0$ main sequence of star forming galaxies (see, e.g., \citealt{Noeske_2007} and \autoref{fig:rings_on_SFMS}). The selected galaxies are massive (stellar mass of $9.5 < \log(M_* / M_\odot) < 11.0$), actively star-forming (specific SFR of $\log({\rm sSFR}/\text{yr}^{-1}) > -11$), relatively nearby ($d \leq 23\,$Mpc), not too edge-on ($i<75\degree$) and have a declination within $-75\degree<\delta<+25\degree$ so that they can be observed by telescopes in the southern hemisphere. For more details on the sample selection, see \citet{Leroy_2021}.

\subsection{PHANGS-ALMA data}\label{sec:CO(2-1)-maps}

The PHANGS-ALMA data set covers observations of the rotational $J=2 \rightarrow 1$ CO emission line, which can be used as a tracer for the most abundant molecule, H$_2$. It contains high-resolution (${\sim} 1\,\arcsec$ corresponding to ${\lesssim} 100\,$pc at $d \lesssim 20\,$Mpc) data cubes of CO(2-1) emission observed with a combination of the ALMA $12\,$m and $7\,$m arrays and the total power antennas, that recover CO emission from all spatial scales. For detailed information about the data reduction and imaging process see \citet{Leroy_2021_reduction}. We use the ``broad mask'' versions of the integrated intensity maps, which achieve high completeness, i.e., including almost all emission, at the expense of a lower S/N.

From the CO(2-1) maps one can obtain the molecular mass of a region $M_\text{mol}^\text{region}$ using the expression
\begin{equation}\label{equ:m_mol}
    M_\text{mol}^\text{region} = \sum_\text{region (px)} \Sigma_\text{mol, px} A_\text{px} = \sum_\text{region (px)} \alpha_\text{CO}^{2-1} I_\text{CO, px}^{2-1} A_\text{px}
\end{equation}
with $\Sigma_\text{mol, px}$ denoting the molecular gas surface density per pixel on the sky, $I_\text{CO, px}^{2-1}$ being the CO(2-1) integrated intensity per pixel, $A_\text{px}$ denoting the pixel size in units of physical area (kpc$^2$), and $\alpha_\text{CO}$ is a conversion factor between CO emission and molecular gas mass with units of a mass-to-light-ratio.
In this study we use maps of $\alpha_\text{CO}^{2-1}$ created by \citet{sun_2025_alphaCO} using the prescription by \citet{schinner_2024}. This prescription attempts to account for excitation variations, emissivity variations, and metallicity effects -- spatially varying both within individual galaxies and between galaxies -- based on synthesizing work by \citet{bolatto_alpha_2013, accurso_17_alpha, Gong_2020, sun_2020_alpha, sun_2022_alpha, denBrok_2021_lines, Hu_2022_alpha, leroy_2022_lines, teng_alpha_CO_2023, chiang_2024_alpha}.
The error on these $\alpha_\text{CO}^{2-1}$ maps are on the order of ${\sim} 0.2{-}0.3$\,dex based on comparisons with observational estimates \citep{sun_2025_alphaCO}.

\begin{table*}[]
    \centering
    \caption{Host galaxy properties.}
    \begin{tabular}{cccccccc}
        \hline
        \hline
        Galaxy & $d$ & $i$ & PA & $M_*^\text{gal}$ & SFR$_\text{gal}$ & $M_\text{mol}^\text{gal}$ & $M_\text{H\,{\sc i}}^\text{gal}$ \\
        & [Mpc] & $[\degree$] & [$\degree$] & [$10^9\, M_\odot$] & [$M_\odot/\text{yr}$] & [$10^9\, M_\odot$] & [$10^9\, M_\odot$]\\
        (1) & (2) & 
        (3) & (4) & 
        (5) & (6) & 
        (7) & (8) \\
        \hline
        NGC\,1097 & 13.6 $\pm$ 2.0  & 48.6 $\pm$ 6.0 & 122.4 $\pm$ 3.6 & 57 $\pm$ 15 & 4.7 $\pm$ 1.2 & 5.52 $\pm$ 0.01 & 4.1 $\pm$ 1.0 \\
        NGC\,1300 & 19.0 $\pm$ 2.9  & 31.8 $\pm$ 6.0 & 278.0 $\pm$ 1.0 & 41 $\pm$ 11 & 1.17 $\pm$ 0.30 & 2.51 $\pm$ 0.02 & 2.4 $\pm$ 0.6 \\
        NGC\,1365 & 19.6 $\pm$ 0.8  & 55.4 $\pm$ 6.0 & 201.1 $\pm$ 7.5 & 98 $\pm$ 25 & 17 $\pm$ 4 & 18.07 $\pm$ 0.05 & 8.7 $\pm$ 2.3 \\
        NGC\,1433 & 18.6 $\pm$ 1.9  & 28.6 $\pm$ 6.0 & 199.7 $\pm$ 0.3 & 73 $\pm$ 19 & 1.13 $\pm$ 0.29 & 1.97 $\pm$ 0.01 & 2.51 $\pm$ 0.7 \\
        NGC\,1512 & 18.8 $\pm$ 1.9  & 42.5 $\pm$ 6.0 & 261.9 $\pm$ 4.2 & 52 $\pm$ 13 & 1.28 $\pm$ 0.33 & 1.33 $\pm$ 0.02 & 7.6 $\pm$ 2.0 \\
        NGC\,1566 & 17.7 $\pm$ 2.0  & 29.5 $\pm$ 10.6 & 214.7 $\pm$ 4.1 & 61 $\pm$ 16 & 4.5 $\pm$ 1.2 & 5.05 $\pm$ 0.01 & 6.4 $\pm$ 1.7 \\
        NGC\,1672 & 19.4 $\pm$ 2.9  & 42.6 $\pm$ 12.9 & 134.3 $\pm$ 0.4 & 54 $\pm$ 14 & 7.6 $\pm$ 2.0 & 7.24 $\pm$ 0.02 & 16.0 $\pm$ 4.2 \\
        NGC\,2566 & 23.4 $\pm$ 3.5  & 48.5 $\pm$ 6.0 & 312.0 $\pm$ 2.0 & 51 $\pm$ 13 & 8.7 $\pm$ 2.3 & 7.17 $\pm$ 0.02 & 2.4 $\pm$ 0.6 \\
        NGC\,2903 & 10.0 $\pm$ 2.5  & 66.8 $\pm$ 3.1 & 203.7 $\pm$ 2.0 & 43 $\pm$ 11 & 3.1 $\pm$ 0.8 & 3.735 $\pm$ 0.005 & 3.4 $\pm$ 0.9 \\
        NGC\,2997 & 14.1 $\pm$ 2.8  & 33.0 $\pm$ 9.0 & 108.1 $\pm$ 0.7 & 54 $\pm$ 14 & 4.4 $\pm$ 1.1 & 6.786 $\pm$ 0.009 & 7.2 $\pm$ 1.9 \\
        NGC\,3351 & 9.96 $\pm$ 0.33 & 45.1 $\pm$ 6.0 & 193.2 $\pm$ 2.0 & 23 $\pm$ 6  & 1.32 $\pm$ 0.34 & 1.23 $\pm$ 0.01 & 0.85 $\pm$ 0.22 \\
        NGC\,3489 & 11.9 $\pm$ 1.6  & 63.7 $\pm$ 5.0 & 70.0 $\pm$ 10.0 & 19 $\pm$ 5  & 0.023 $\pm$ 0.006 & 0.048 $\pm$ 0.001 & 0.025 $\pm$ 0.007 \\
        NGC\,4303 & 17.0 $\pm$ 3.0  & 23.5 $\pm$ 9.2 & 312.4 $\pm$ 2.5 & 33 $\pm$ 9  & 5.3 $\pm$ 1.4 & 8.12 $\pm$ 0.02 & 4.6 $\pm$ 1.2 \\
        NGC\,4321 & 15.2 $\pm$ 0.5  & 38.5 $\pm$ 2.4 & 156.2 $\pm$ 1.7 & 56 $\pm$ 14 & 3.6 $\pm$ 0.9 & 7.77 $\pm$ 0.02 & 2.7 $\pm$ 0.7 \\
        NGC\,4459 & 15.9 $\pm$ 2.2  & 47.0 $\pm$ 5.0 & 108.75 $\pm$ 10.0 & 48 $\pm$ 12 & 0.22 $\pm$ 0.06 & 0.264 $\pm$ 0.003 & - \\
        NGC\,4476 & 17.5 $\pm$ 2.4  & 60.1 $\pm$ 5.0 & 27.4 $\pm$ 10.0 & 6.5 $\pm$ 1.7 & 0.04 $\pm$ 0.01 & 0.071 $\pm$ 0.002 & - \\
        NGC\,4477 & 15.8 $\pm$ 2.4  & 33.5 $\pm$ 5.0 & 25.7 $\pm$ 10.0 & 39 $\pm$ 10 & 0.08 $\pm$ 0.02 & 0.038 $\pm$ 0.002 & - \\
        NGC\,5236 & 4.90 $\pm$ 0.18 & 24.0 $\pm$ 5.0 & 225.0 $\pm$ 10.0 & 34 $\pm$ 9 & 4.2 $\pm$ 1.1 & 2.8 $\pm$ 0.2 & 9.5 $\pm$ 2.5 \\
        NGC\,5248 & 14.9 $\pm$ 1.3  & 47.4 $\pm$ 16.3 & 109.2 $\pm$ 3.5 & 25 $\pm$ 7 & 2.3 $\pm$ 0.6 & 4.541 $\pm$ 0.007 & 3.2 $\pm$ 0.8 \\
        NGC\,7743 & 20.3 $\pm$ 2.8  & 37.1 $\pm$ 5.0 & 86.2 $\pm$ 10.0 & 23 $\pm$ 6 & 0.21 $\pm$ 0.06 & 0.360 $\pm$ 0.005 & 0.31 $\pm$ 0.08 \\
        \hline
    \end{tabular}
    \label{table:gal-props}
    \tablefoot{{The host properties were compiled by \citet{Leroy_2021}}: galaxy distance and its error based on the distance determination method from \citet{anand_2021} in Column (2); the inclination and position angle in Cols. (3), (4) (\citealt{Lang_2020} or \citealt{makarov_hyperleda_2014}); the stellar mass and its error in Col. (5) \citep{Leroy_2021}; the SFR and its error in Col. (6)  \citep{Calzetti_2007, Leroy_2019, Leroy_2021, belfiore_2023}{}{}; the molecular gas mass and its error in Col. (7) (These values are taken from the PHANGS sample table released together with the ALMA data products described in \citealt{Leroy_2021}. They include aperture corrections and adopt a varying metallicity-dependent $\alpha_\text{CO}$. The uncertainties include statistical errors only, thus appearing smaller than the conservatively estimated errors on the ring molecular gas mass.); the atomic gas mass and its error in Col. (8) \citep{Leroy_2021}. ``-'' means that no data for the H\,{\sc i} mass was available from \citet{Leroy_2021}.}
\end{table*}

\subsection{PHANGS-MUSE -- $\Sigma_\text{SFR}$ maps}\label{sec:SFR-maps}

For 51 PHANGS galaxies -- including 14 with central rings -- maps of star formation rate surface density ($\Sigma_\text{SFR}$) are available. They are derived from H$\alpha$ emission from MUSE archival data and the sample from \citet{belfiore_2023}, and their resolution reaches the common optimum (copt) resolution of MUSE (${\sim}(0.95\pm 0.16)\,\arcsec$, depending on the observing conditions, \citealt{emsellem_muse_2022}). The copt resolution is obtained by convolving the data at all wavelengths and from all pointings within a mosaic to a common angular resolution image of a galaxy. All maps are derived by correcting the observed H$\alpha$ emission for dust attenuation based on the Balmer decrement and then converting it into $\Sigma_\text{SFR}$ maps using the conversion factor of \citet{Calzetti_2007} \citep[for details, see][]{belfiore_2023}:
\begin{equation}\label{equ:sfr_calzetti}
    \text{SFR}(M_{\odot} \, \text{yr}^{-1}) = 5.5 \cdot 10 ^{-42}L(\text{H}\alpha)_\text{corr} (\text{ergs s}^{-1}).
\end{equation}

\noindent This conversion factor was computed with Starburst99 \citep{leitherer_1999_starburst99}, assuming a constant star formation history, an age of $100\,$Myr, solar metallicity, and a \citet{Kroupa_2001_IMF} initial mass function (IMF).

\subsection{Host galaxy and bar properties}\label{sec:sample-tab}

The host galaxy properties were derived and compiled by \citet{Leroy_2021} and are provided in \autoref{table:gal-props}. For 15 of the 20 galaxies with central rings, the deprojected bar semi-major axis as a measure for the bar length $L_\text{bar}$ and the bar ellipticity $\varepsilon_\text{bar}$ were measured by \citet{herrera-endoqui_catalogue_2015}, while the bar maximum gravitational torque $Q_\text{b}$ and the bar maximum normalized 
\begin{table*}[!ht]
    \centering
    \caption{Central ring properties.}
    \begin{tabular}{ccccccc}
        \hline
        \hline
        Galaxy & $r_\text{ring}$ & $L_\text{CO}^\text{ring}$  & $M_\text{mol}^\text{ring}$ &  ${M_\text{mol}^\text{ring}}/{M_\text{mol}^\text{gal}}$ & SFR$_\text{ring}$ & ${\text{SFR}_\text{ring}}/{\text{SFR}_\text{gal}}$ \\
        &  [pc] & [K km/s pc$^2$] & [$10^6\,M_\odot$] & [\%] & [$M_\odot/\text{yr}$] & [\%] \\
        (1) & (2) & (3) & (4) & (5) & (6) & (7) \\
        \hline
        NGC\,1097 & 970 $\pm$ 230 & 300 $\pm$ 50 & 550 $\pm$ 180 & 10.0 $\pm$ 3.3 & 1.9 $\pm$ 0.6 & 40 $\pm$ 17 \\
        NGC\,1300 & 400 $\pm$ 110 & 77 $\pm$ 16 & 140 $\pm$ 50 & 5.5 $\pm$ 1.9 & 0.102 $\pm$ 0.032 & 8.7 $\pm$ 3.5 \\
        NGC\,1365 & 830 $\pm$ 210 & 1500 $\pm$ 340 & 1900 $\pm$ 450 & 10.5 $\pm$ 2.5 & 6.3 $\pm$ 1.4 & 37 $\pm$ 13 \\
        NGC\,1433 & 1070 $\pm$ 150 & 94 $\pm$ 32 & 170 $\pm$ 60 & 8.5 $\pm$ 3.2 & 0.038 $\pm$ 0.026 & 3.3 $\pm$ 2.4 \\
        NGC\,1512 & 710 $\pm$ 150 & 80 $\pm$ 10 & 137 $\pm$ 31 & 10.3 $\pm$ 2.4 & 0.24 $\pm$ 0.06 & 19 $\pm$ 7 \\
        NGC\,1566 & 840 $\pm$ 190 & 71 $\pm$ 7 & 100 $\pm$ 24 & 2.0 $\pm$ 0.5 & 0.052 $\pm$ 0.016 & 1.1 $\pm$ 0.5 \\
        NGC\,1672 & 630 $\pm$ 170 & 151 $\pm$ 33 & 540 $\pm$ 190 & 7.4 $\pm$ 2.7 & 2.9 $\pm$ 1.1 & 38 $\pm$ 17 \\
        NGC\,2566 & 300 $\pm$ 80 & 220 $\pm$ 90 & 400 $\pm$ 200 & 5.6 $\pm$ 2.7 & - & - \\
        NGC\,2903 & 260 $\pm$ 90 & 270 $\pm$ 70 & 120 $\pm$ 50 & 3.3 $\pm$ 1.5 & 0.64 $\pm$ 0.29 & 21 $\pm$ 11 \\
        NGC\,2997 & 410 $\pm$ 120 & 160 $\pm$ 40 & 140 $\pm$ 60 & 2.0 $\pm$ 0.9 & - & - \\
        NGC\,3351 & 370 $\pm$ 70 & 116 $\pm$ 21 & 89 $\pm$ 17 & 7.2 $\pm$ 1.4 & 0.31 $\pm$ 0.05 & 24 $\pm$ 7 \\
        NGC\,3489 & 152 $\pm$ 33 & 5.0 $\pm$ 1.9 & 3.2 $\pm$ 1.5 & 7 $\pm$ 3 & 0.014 $\pm$ 0.008 & 61 $\pm$ 38 \\
        NGC\,4303 & 290 $\pm$ 100 & 56 $\pm$ 22 & 120 $\pm$ 60 & 1.5 $\pm$ 0.8 & 0.24 $\pm$ 0.10 & 4.4 $\pm$ 2.1 \\
        NGC\,4321 & 560 $\pm$ 120 & 160 $\pm$ 50 & 270 $\pm$ 90 & 3.5 $\pm$ 1.1 & 0.51 $\pm$ 0.13 & 14 $\pm$ 5 \\
        NGC\,4459 & 120 $\pm$ 50 & 23 $\pm$ 4 & 13 $\pm$ 4 & 5.0 $\pm$ 1.5 & - & - \\
        NGC\,4476 & 460 $\pm$ 120 & 34 $\pm$ 4 & 43 $\pm$ 12 & 60 $\pm$ 18 & - & - \\
        NGC\,4477 & 120 $\pm$ 40 & 15 $\pm$ 4 & 9.8 $\pm$ 3.6 & 25 $\pm$ 10 & - & - \\
        NGC\,5236 & 230 $\pm$ 40 & 430 $\pm$ 100 & 101 $\pm$ 24 & 3.5 $\pm$ 0.9 & - & - \\
        NGC\,5248 & 550 $\pm$ 120 & 130 $\pm$ 36 & 130 $\pm$ 40 & 2.9 $\pm$ 1.0 & 0.31 $\pm$ 0.13 & 14 $\pm$ 7 \\
        NGC\,7743 & 130 $\pm$ 50 & 39 $\pm$ 8 & 36 $\pm$ 12 & 10.1 $\pm$ 3.4 & 0.062 $\pm$ 0.026 & 29 $\pm$ 14 \\
        MW CMZ\protect\footnotemark & 150 $\pm$ 50 & - & 30$^{+20}_{-10}$ & 4.6$^{+3.1}_{-1.5}$ & 0.07$^{+0.08}_{-0.02}$ & 3.7$^{+4}_{-1.3}$\\
        \hline
    \end{tabular}
    \label{table:ring-props}
    \tablefoot{The central ring radius and its errors in Col.~(2); The central ring CO(2-1) Luminosity and its error in Col.~(3); the central ring molecular gas mass and its error in Col.~(4); the central ring molecular gas fraction and its error in Col.~(5); the central ring SFR and its error in Col.~(6); the central ring SFR fraction and its error in Col.~(7). ``-'' means that the central ring SFR (and therefore, also its SFR fraction) could not be determined due to the fact that $\Sigma_\text{SFR}$ maps were only available for $14$ of the $20$ galaxies. Literature values for the MW CMZ are taken from \citet{Henshaw_2023_CMZ, mills2017milky}.}
\end{table*}
\footnotetext{Values obtained from \citet{Henshaw_2023_CMZ, mills2017milky}, or calculated with global MW values from \citet{Chomiuk_2011, Roman-Duval_2016}.}
$m=2$ Fourier amplitude $A_2^\text{max}$ were determined by \citet{diaz-garcia_characterization_2016}.

\section{Central ring identification and measurements}\label{sec:analysis}

\subsection{Sample selection}\label{sec:samp-select}

We visually selected a sample of central rings from the original PHANGS-ALMA sample \citep{Leroy_2021} using the following criteria: (a) The available moment-0 map must have an angular resolution of ${\sim}1\,\arcsec$, corresponding to ${\lesssim} 100\,$pc at distances $\lesssim 20\,$Mpc, such that central rings with a radius of $r > 100\,$pc can in principle be resolved, which reduces our parent sample to $81$ PHANGS-ALMA galaxies. (b) In the molecular gas distribution a central ring-like structure must be visible with a clearly identifiable shape and size. This means, we must be able to resolve substructure within the ring-region, i.e. either a hole in the center (e.g. NGC\,4476, NGC\,4477, see Figure \ref{fig:rings_gallery}), or a central emission component clearly distinct from the ring around it (e.g. NGC\,1097, NGC\,1512). This excludes galaxies where we only see a central gas concentration and galaxies with diffuse and/or clumpy central emission where we cannot reliably define a ring-like distribution. As central rings can have different geometries \citep[e.g.,][]{jimenez-sanchez_2025_rings}, we consider the following geometries: (i) ``classical'' circum-nuclear ring within a bar, (ii) tightly wound spiral arms forming a pseudo-ring, and (iii) other types of rings. We also do not apply a morphological cut for the host galaxies and include central rings in both spiral and early type galaxies that are part of the PHANGS-ALMA sample. See also \autoref{app:ring_sample} for comments on the individual rings in our sample.

\citet{stuber_gas_2023} already carried out a first classification of different structures in the molecular gas distribution of the PHANGS-ALMA sample, including central rings.
We compare their classification based on the two before-mentioned criteria and confirm $20$ out of $22$ central rings reported by \citet{stuber_gas_2023}. The two central rings in Circinus and NGC\,1068 listed in \citet{stuber_gas_2023} do not fulfill our criterion for the data resolution, as these galaxies were observed with ALMA's $7\,$m array and total power antennas only, and are therefore not considered in this work.

Maps of $\Sigma_\text{SFR}$ are available for $14$ of these $20$ galaxies (see Sect.\,\ref{sec:SFR-maps}), and are used to derive SFRs and depletion times $t_\text{dep} = M_\text{mol}/\text{SFR}$ in the central rings. In \autoref{table:ring-props} the identified central rings along with their measured properties are provided. In \autoref{fig:rings_on_SFMS}, all galaxies in this study are shown at their location with respect to the main sequence of star-forming galaxies along with the full PHANGS galaxy sample. Almost all our central rings are found in rather massive galaxies ($\log(M_* / M_\odot)>10.1$), which is in agreement with observations by \citet{fraser-mckelvie_2020_SFR-bar} and simulations carried out by \citet{verwilghen_2024} (see Sect.\,\ref{sec:disc:SFMS-stellar_mass} and Emsellem, in prep.).

\subsection{Determination of ring geometry}\label{sec:ring-shapes}

\begin{figure}[t!]
    \centering
    \includegraphics[width=0.5\textwidth]{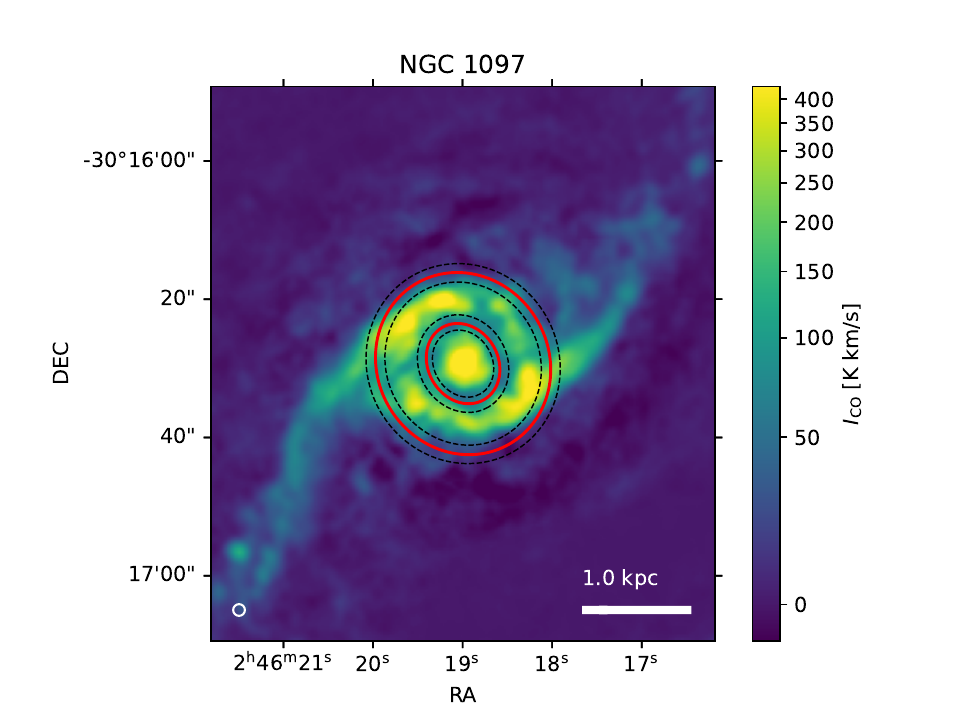}
    \caption{Molecular gas distribution of the galaxy NGC\,1097. The different ellipses represent the ``best'' elliptical annulus (red, solid line) and uncertainty ellipses (black, dashed lines), indicating emission which is definitely ring material (narrower annulus, ``strict'' mask) and emission that might still be ring material (wider annulus, ``broad'' mask). The white circle in the lower left corner indicates the beam size, and a scalebar is provided in the lower right. The color bar indicates the integrated intensity in terms of the brightness temperature.}
    \label{fig:NGC1097_annuli}
\end{figure}

We used the SAOImageDS9\footnote{\url{https://sites.google.com/cfa.harvard.edu/saoimageds9/about}} tool (hereafter DS9), to visually determine elliptical annuli which best describe the shape of central rings in the molecular gas distribution as traced by CO(2-1). We use the projected images of the galaxies to avoid introducing uncertainties by deprojection.

\begin{figure*}
    \centering
    \includegraphics[width=0.8\textwidth]{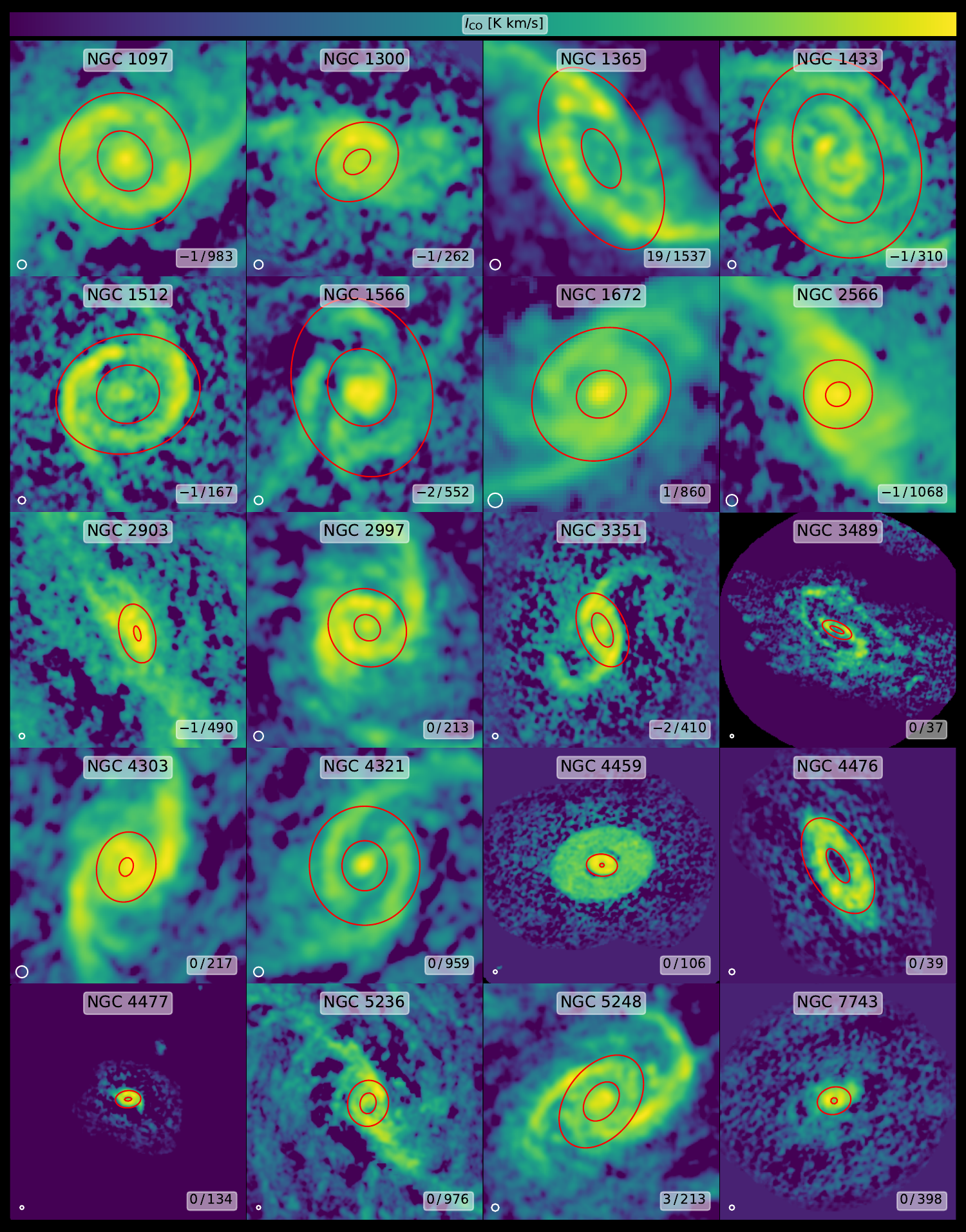}
    \caption{Gallery of central rings identified in a sample of 81 PHANGS-ALMA galaxies. The galaxy name is given at the top of each panel. All panels have the same physical size of $3 \times 3\,\text{kpc}^2$, but with differing color stretches. The minimum and maximum intensity is listed at the bottom right of each panel. The ``best'' elliptical annulus (see Sect.\,\ref{sec:ring-shapes}) is plotted in red and the beam size is plotted in white in the lower left corner.}
    \label{fig:rings_gallery}
\end{figure*}

An elliptical annulus can be described by seven parameters: major and minor axes of both the inner and outer ellipse ($h_\text{in/out}$ and $w_\text{in/out}$), position of the center $(x_\text{c},y_\text{c})$ of the ellipses, and their position angle PA. We allowed the inner and outer ellipse of the ring region to have different ellipticities $\varepsilon = 1-w/h$, as the inner and outer boundary of the central rings can have different shapes (see \autoref{fig:rings_gallery}). As default, we used the host galaxy centers as the center for the ellipses. For $12/20$ central rings, we needed to slightly shift (mostly $0.5{-}1.5\,\arcsec$) the center to correctly capture the molecular gas distribution.

Using different color stretches in DS9, we determined heights, widths, and PAs that best describe the central rings (``best'' ellipses, see \autoref{fig:NGC1097_annuli}). Similary, uncertainty ellipses for the ``best'' rings were identified by defining an ellipse that contains all the molecular gas emission which might be still ring material (``broad'' uncertainty ellipses), and a second one that only captures emission that is definitely in the ring (``strict'' uncertainty ellipses). As the resolution of the ALMA data is set by the beam size (see \citealt{Leroy_2021_reduction}), we define the uncertainty ellipses to differ from the nominal best ring at least by $0.5$ times the beam size. The PAs for the uncertainty annuli are the same as for the ``best'' ones, as varying the PA introduces only a small error compared to the one introduced by varying the annulus width. The determined ellipse parameters for the rings are presented in \autoref{app:ring-masks}.

The central ring radius $r_\text{ring}$ is the deprojected mean of the outer and inner ``best'' semi-major axes, which we also compile in \autoref{table:ring-props}.

\subsection{Central ring molecular gas mass}\label{sec:ring-masses}

We use the integrated quantities ($M_\text{mol}^\text{ring}$, SFR$_\text{ring}$) as indicators of the ring properties as they are more robust than surface densities of molecular gas and star formation against the exact determination of the ring shape and presence of regions of low surface density such as, e.g., the ring edges. In particular, the ring area in the galaxy plane is sensitive to the inclination and its uncertainty, which we wanted to avoid.

We obtain the molecular gas mass of the central ring with \autoref{equ:m_mol}, when applied to the area of the ring (elliptical annulus, see Sect.\,\ref{sec:ring-shapes}).
The error on the central ring molecular gas mass is dominated by the uncertainty from varying the ring shape, i.e. from integrating the ``broad'' and ``strict'' uncertainty annuli. Taking the mean of the two provides a first component of the uncertainty, $\Delta M_\text{mol, size}^\text{ring}$. We use noise maps $\Delta I_\text{CO, px}^{2-1}$ for the CO(2-1) moment-0 maps \citep{Leroy_2021_reduction} to obtain a second uncertainty component based on standard Gaussian error propagation:
\begin{equation}\label{equ:delta_M}
    \Delta M_\text{mol, noise}^\text{ring} = A_\text{px} \cdot \sqrt{\sum_\text{ring (px)} \bigl(\alpha_\text{CO}^{2-1} \Delta I_\text{CO, px}^{2-1}\bigr)^2}.
\end{equation}
The uncertainty on the distance affects the pixel size $A_\text{px}$ and thus the central ring molecular gas mass, giving the third uncertainty component, $\Delta M_\text{mol, dist}^\text{ring}$. The three uncertainty components are then added in quadrature yielding the total uncertainty $\Delta M_\text{mol}^\text{ring}$.
The $\alpha_\text{CO}^{2-1}$ determination has an uncertainty of ${\sim}0.2{-}0.3\,$dex \citep{sun_2025_alphaCO} depending on the exact prescription used. Given that this is a systematic bias, we do not include it in our error calculations here, and separately discuss the effect of using different prescriptions for $\alpha_\text{CO}$ in Sect.\,\ref{sec:disc:bars-funneling}.

\subsection{Central ring SFRs}\label{sec:ring-SFRs}

For the $14$ galaxies of our ring sample with SFR maps (see Sect.\,\ref{sec:SFR-maps}), the central ring SFR is measured analogous to the molecular gas mass. We apply the same elliptical annular masks to the $\Sigma_\text{SFR}$ maps and integrate the SFR surface density over the area of the central rings. The uncertainty determination is similar to the one in the previous section: The uncertainty on the central ring shape applies here in the same way, the $\Sigma_\text{SFR}$ maps also have associated error maps that must be taken into account, and the distance error also affects the area.

\subsection{Central ring sample}\label{sec:ring-comments}

Central rings are formed easily in simulations of barred galaxies \citep[e.g.,][and many others]{athanassoula_1992, kim_2012_rings, sormani_2024} and observations show that they are found preferentially in barred galaxies \citep{knapen_2005, comeron_ainur_2010, stuber_gas_2023}. Interestingly, both \citet{stuber_gas_2023} and this study identify small ring-like features in the centers of the three non-barred galaxies NGC\,2997\footnote{The bar classification of NGC\,2997 is ambiguous though, see \autoref{app:ring_sample}.}, NGC\,4459, and NGC\,4476. These central rings might have formed by a different mechanism than the classical central ring induced by a non-axisymmetric potential and thus might not be directly comparable with the latter. Furthermore, as we here investigate all ring-like central structures within the PHANGS sample (see Sect.\,\ref{sec:samp-select}), we might have included also so-called pseudo-rings, that are made up of tightly wound spiral arms. These rings might not be directly comparable with the ``classical'' central ring. In total, five of the central rings we identified are found in early type galaxies (NGC\,3489, NGC\,4459, NGC\,4476, NGC\,4477, NGC\,7743, see Sect.\,\ref{sec:samp-select} and \autoref{app:ring_sample}), and their gas dynamics might differ from those in late-type spiral galaxies,
given the larger importance of the stronger stellar potential relative to gas self-gravity \citep[e.g.][]{Davis_2013_gasETG, meidt_2018_gasDyn, liu_2021_gasDyn}, which tends to lead to a smoother gas structure and morphology \citep{davis_2022_gasDyn}.
Therefore, the rings found in ETGs are highlighted throughout the following analysis.

The molecular gas distribution of all central rings identified in the PHANGS-ALMA sample often exhibits bar lanes for the barred galaxies (see Fig.\,\ref{fig:rings_gallery}). Furthermore, we comment on each individual central ring in \autoref{app:ring_sample}.


\begin{figure*}
    \centering
    \includegraphics[width=0.9\textwidth]{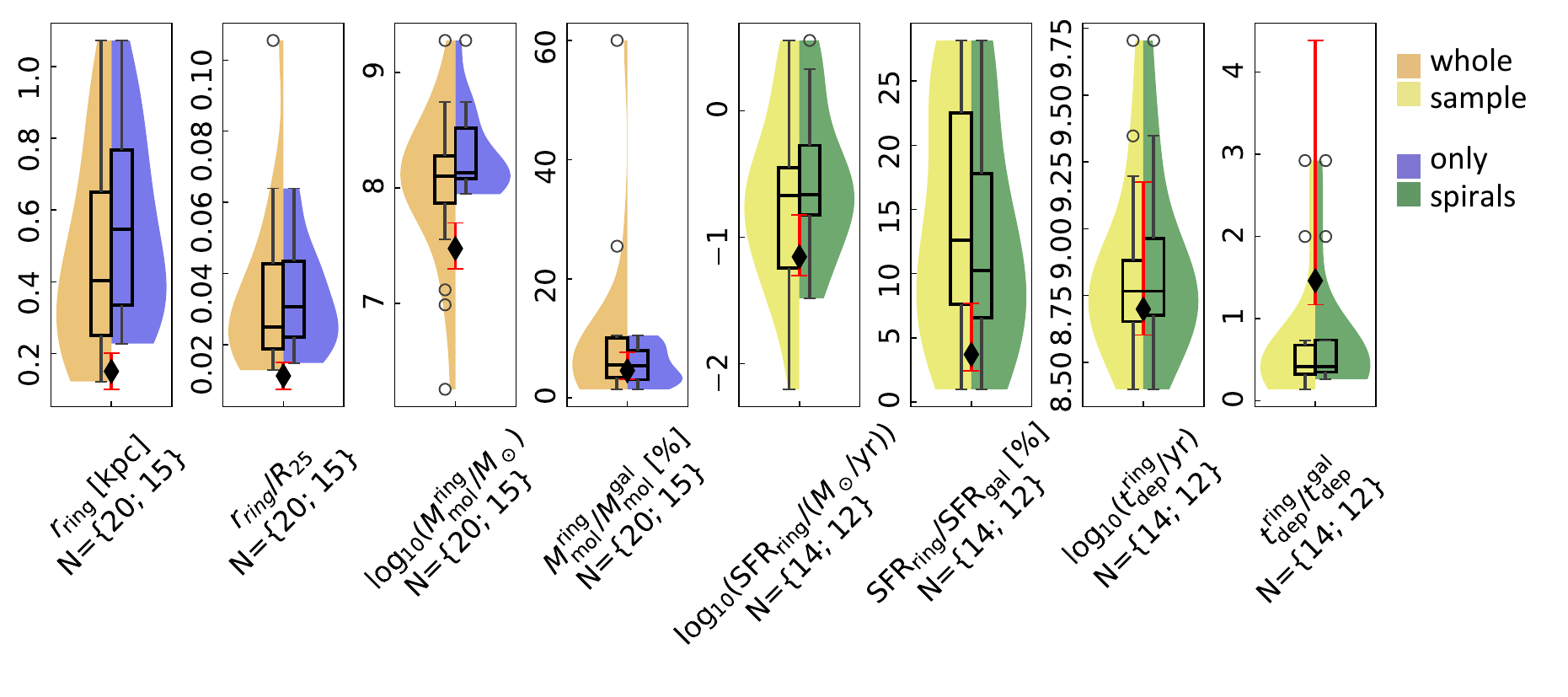}
    \caption{Distributions of central ring parameters (from left to right): radius $r_\text{ring}$, relative radius $r_\text{ring}/r_{25}$, molecular gas mass $M_\text{mol}^\text{ring}$, ring molecular gas mass fraction $M_\text{mol}^\text{ring}/ M_\text{mol}^\text{gal}$, star formation rate ${\rm  SFR}_\text{ring}$, ring SFR fraction ${\rm SFR}_\text{ring} / {\rm SFR}_\text{gal}$, depletion time $M_\text{mol}^\text{ring} / {\rm SFR}_\text{ring}$, and relative depletion time $t_\text{dep}^\text{ring} / t_\text{dep}^\text{gal}$. The distributions are shown for the whole sample on the left and for spirals only on the right, thus leading to the different numbers $N= \{...\}$ considered for the distributions. The violins show a smoothed distribution, the boxes contain the $25$th to $75$th percentile of the distributions, the black whiskers span out to $1.5$ times the interquartile range, and the circles mark outliers beyond this range. The median is shown by the black horizontal line. The respective values for the MW CMZ are plotted as black diamonds along with their error bars in red.}
    \label{fig:violin-distr}
\end{figure*}

\begin{table*}[]
    \centering
    \caption{Central ring properties - medians and ranges}
    \begin{tabular}{|c|c|c|c|c|c|c|}
        quantity & unit & \multicolumn{2}{|c|}{median} & \multicolumn{2}{|c|}{interquartile range} & MW CMZ \\
         & & all & Spirals & all & Spirals &  \\
         \hline
         $r_\text{ring}$ & pc & 400 & 550 & $-150, +250$ & $-210, +220$ & 150 $\pm$ 50 \\
         $\log_{10}( M_\text{mol}^\text{ring})$ & $M_\odot$ & 8.1 & 8.14 & $-0.23, +0.17$ & $-0.06, +0.4$ & $7.5\pm0.2$ \\
         $M_\text{mol}^\text{ring}/M_\text{mol}^\text{gal}$ & \% & 5.6 & 5.5 & $-2.1, +4.5$ & $-2.4, +2.5$ & 4.6$^{+3.1}_{-1.5}$ \\
         $\log_{10}( L_\text{CO(2-1)}^\text{ring})$ & K km/s pc$^2$ & 5.02 & 5.18 & $-0.31, +0.22$ & $-0.24, +0.21$ & - \\
         $\text{SFR}_{\text{ring}}$ & M$_\odot$/yr & 0.21 & 0.22 & $-0.16, +0.15$ & $-0.07, +0.40$ & 0.07$^{+0.08}_{-0.02}$ \\
         $\text{SFR}_\text{ring} / \text{SFR}_\text{gal}$ & \% & 13 & 10.2 & $-5, +10$ & $-3.7, +8$ & 3.7$^{+4}_{-1.3}$ \\
         $t_\text{dep}^\text{ring}$ & Gyr & 0.58 & 0.58 & $-0.14, +0.18$ & $-0.11, +0.39$ & $0.5_{-0.1}^{+1.0}$
    \end{tabular}
    \label{table:ring-props_2}
\end{table*}

\section{Results}\label{sec:results}

We present the results on central ring properties and compare them with literature values for the MW CMZ. This will not only improve our understanding of central rings in different environments but the interpretation of the MW CMZ data in the context of nearby normal star-forming galaxies. We further compare our ring fractions and radii to literature findings, in particular the largest literature sample of central rings, the ``Atlas of Images of Nuclear Rings'' (AINUR, \citealt{comeron_ainur_2010}), which used UV, H$\alpha$, and Pa$\alpha$ as tracers for star-forming rings, as well as color-index and structure maps \citep[][]{pogge_2002_structure-map} for detecting dust rings. Furthermore, we test for expected correlations between the central ring molecular gas mass and literature bar parameters.

\subsection{PHANGS central ring properties}\label{sec:ring-props}

In \autoref{table:ring-props_2} we list properties of the central rings across our sample. We give median values and ranges for both the whole sample and the sample without the ETGs (spirals only), as we later compare the MW CMZ only with the central rings found in spiral galaxies. The interquartile ranges in the table are given as uncertainty values on the median. The distributions of these central ring properties along with the respective values for the MW CMZ are also presented as violin plots in \autoref{fig:violin-distr}.

The radii of PHANGS central rings are obtained by deprojecting the mean of the outer and the inner ``best'' semi-major axes. Due to the resolution limit, we might have missed central rings with radii $\lesssim100\,$pc,  (see Sect.\,\ref{sec:CO(2-1)-maps}). Upcoming work by Neumann (in prep.) and Emsellem (in prep.) will probe a possible population of central rings with $r\lesssim100\,$pc using higher resolution JWST observations.

The central rings in PHANGS have molecular gas masses of typically ${\sim} 10^{8.1}\,M_\odot$ and contain ${\sim} 1.5{-}10.5\,\%$ of the total molecular gas mass of the galaxy, except for NGC\,4476 (with $(60 \pm 17)\,\%$) and NGC\,4477 (with $(25 \pm 10)\,\%$). Both are ETGs and are expected to behave differently than spiral galaxies \citep{Davis_2013_gasETG, davis_2022_gasDyn, williams_2025_ETG}.

The $14$ central rings with high resolution $\Sigma_\text{SFR}$ maps from MUSE have SFRs of ${\sim} 0.21\,M_\odot / \text{yr}$, contributing ${\sim} 1{-}25\,\%$ to their host galaxy's SFR. We checked whether the sample with available SFR maps is representative of the whole ring sample by comparing the radius and molecular gas mass distributions of the respective (sub-)samples and find that the central rings with SFR maps have slightly larger radii ($550\pm240\,$pc instead of $400^{+250}_{-150}\,$pc), while the overall radius and molecular gas mass distributions do not change much (see \autoref{app:further_biases}).

\subsection{MW CMZ properties}\label{sec:CMZ-props}

The central ${\sim} 1\,$kpc of the Milky Way (MW) hosts several different structures identified in its molecular gas distribution, with the Central Molecular Zone (CMZ) being a ring-like structure that is similar to extragalactic central rings \citep[see review by][]{Henshaw_2023_CMZ}. As the MW is a barred spiral galaxy, we compare the MW CMZ properties with properties of PHANGS central rings that reside within spiral galaxies.

The CMZ has a radius of ${\sim} 100 {-} 200\,$pc\footnote{Where the uncertainty is dominated by our view through the galactic disk.} \citep{Henshaw_2023_CMZ} and is therefore smaller than PHANGS central rings in spiral galaxies. However, there might be a population of rings within this radius range in PHANGS-ALMA, that could not be detected due to the resolution limit\footnote{Because we need to resolve substructure within the ring to identify it as such, see criterion (b) in Sect.\,\ref{sec:samp-select}.}. Thus, the MW CMZ might turn out to be within the distribution if higher resolution data were available.

The MW CMZ has a molecular gas mass of $3_{-1}^{+2} \times 10^7\,M_\odot \approx 10^{7.47}\,M_\odot$ \citep{mills2017milky}, and using $6.5\times10^8\,M_\odot$ \citep{Roman-Duval_2016} for the total molecular gas mass of the Milky Way, its molecular gas mass fraction is $4.6_{-1.5}^{+3.1}\,\%$ \citep[comparable to ${\sim} 5\,\%$ in][]{mills2017milky}. Therefore, while its mass is lower than that of PHANGS central rings in spirals, it contains a similar fraction of the molecular gas of the host galaxy as our ring sample.

The CMZ has an SFR of $0.07^{+0.08}_{-0.02}\,M_\odot/\text{yr}$ \citep{Henshaw_2023_CMZ}. Comparing this value to the SFR of the entire MW of $(1.9\pm0.4)\,M_\odot/\text{yr}$ \citep{Chomiuk_2011}, this corresponds to an SFR fraction of $3.7_{-1.3}^{+4.0}\,$\%. So the MW CMZ's SFR and SFR fraction is on the low-end side -- but still within -- the PHANGS central ring SFR distribution.

The depletion time of the MW CMZ is $t_\text{dep}^\text{CMZ} = 0.5_{-0.1}^{+1.0}\,$Gyr \citep{Henshaw_2023_CMZ}. This is well within the range of depletion times derived for PHANGS central rings.

\subsection{Central ring fraction and radii -- comparison to AINUR}\label{sec:radii_ainur}

\begin{figure}[t!]
    \centering
    \includegraphics[width=0.42\textwidth]{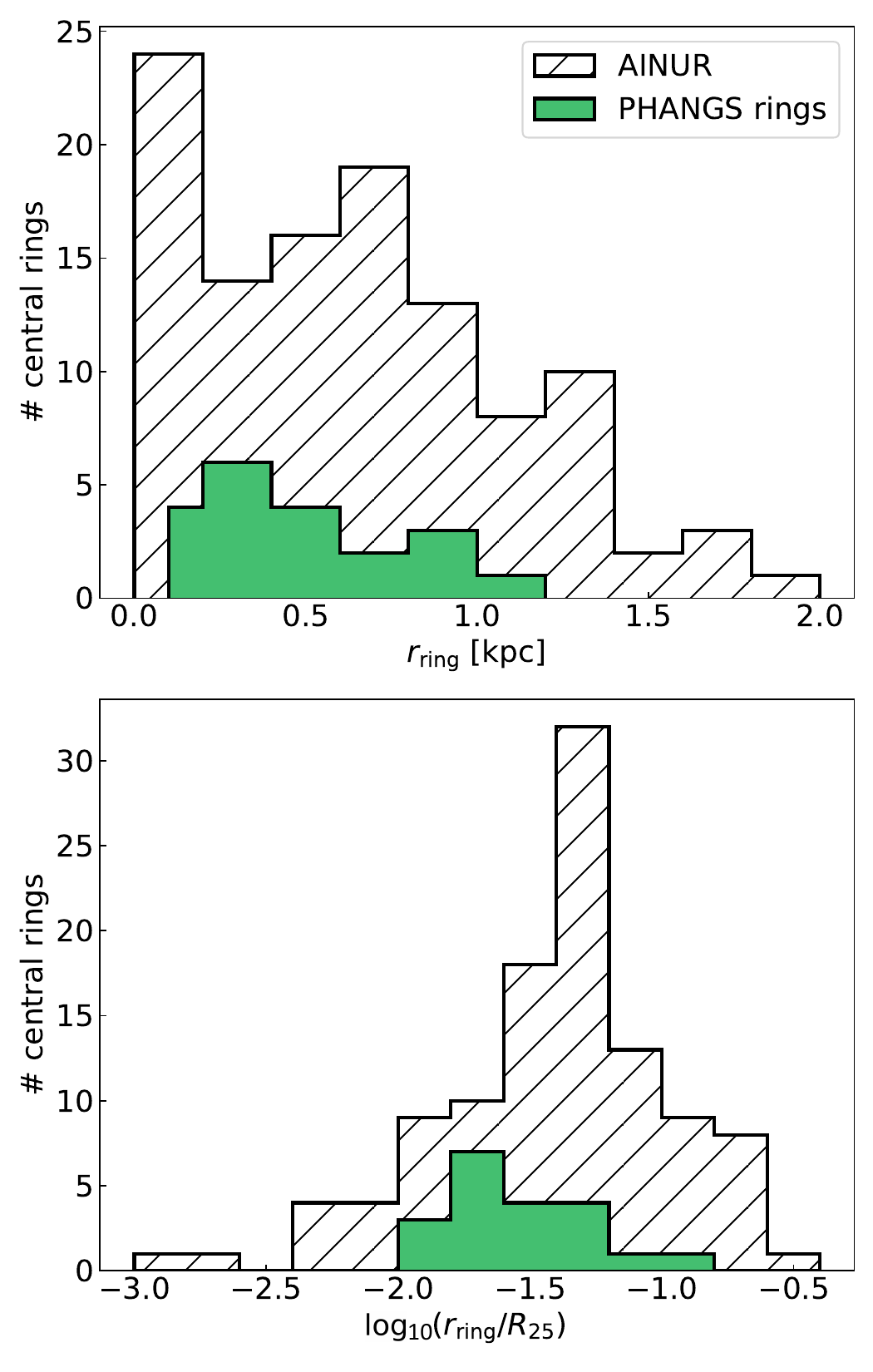}
    \caption{Distribution of the absolute central ring radius (upper panel) and the ring radius relative to the disk size as measured by $r_{25}$ (lower panel). The green histogram shows the sample of central rings in this study while the white histogram shows the ring sample of the ``Atlas of Images of NUclear Rings'' (AINUR; \citealt{comeron_ainur_2010}).}
    \label{fig:r_ring_distr}
\end{figure}

AINUR \citep{comeron_ainur_2010} is a comprehensive study of $113$ central rings in $107$ galaxies as traced by UV, H$\alpha$, Pa$\alpha$, color-index, and structure maps (derived from HST data). In the following we compare our results to the literature findings, in particular from AINUR.

\citet{comeron_ainur_2010} report that $(20 \pm 2)\,\%$ of disk galaxies (with morphological types between $T = -3$ and $T = 7$, similar to PHANGS morphological types between  $T = -2.9$ and $T = 9$) harbor central rings, which is similar to our central ring fraction of $(25 \pm 4)\,\%$. Also \citet{knapen_2005} determines a similar fraction of $(21\pm5)\,\%$ of disk galaxies to host central rings when using ground-based, narrowband H$\alpha$ as tracer, and \citet{erwin_2024_nuc_rings} finds a fraction of $(20\pm3)\,\%$ of barred S0 -Sd galaxies to host central rings. Using the same sample as this study, \citet{stuber_gas_2023} report a central ring fraction of $(31 \pm 4)\,\%$ (and $(45\pm7)\,\%$ in barred galaxies), where the slightly higher number comes from their selection criteria: They pre-selected the galaxies based on the visibility of the disk in molecular line emission, obtaining $72$ valid galaxies for the classification.

A closer comparison with AINUR shows that most molecular central rings have a direct counterpart seen in Pa$\alpha$ or H$\alpha$ emission, with only a few central rings where both studies disagree:
\citet{comeron_ainur_2010} report central rings for the PHANGS galaxies NGC\,1068, NGC\,1317, NGC\,4571, NGC\,4579, and NGC\,4826, but we did not include these in our sample, either because the resolution was too poor (NGC\,1068; see \autoref{sec:samp-select}), or no molecular central ring was identified (NGC\,1317, NGC\,4571), the ring shape was unclear (NGC\,4579), or only unresolved emission was present in the center, which might be resolved as a ring with higher-resolution imaging (NGC\,4826, $r_\text{AINUR} = 150\,$pc).

On the other hand, our ring sample includes NGC\,1365, NGC\,2566, NGC\,3498, NGC\,4476, NGC\,4477, and NGC\,7743, that are not reported by \citet{comeron_ainur_2010}. This is because the ring is obscured by dust (NGC\,1365), they considered it as a pseudo-ring (NGC\,2566), the host galaxy was too inclined (NGC\,3498), no HST data were available (NGC\,4476, NGC\,7743), or no obvious ring in the HST images was detected (NGC\,4477, Comerón, priv. com.). For all other 71 galaxies in the PHANGS survey, especially the remaining 14 with central rings, AINUR and this study agree.

The ring radii in AINUR span a range of ${\sim} 50 {-} 2000\,$pc with more rings at the low-end of the distribution \citep[see Figure 4 in][]{comeron_ainur_2010}. As AINUR made use of HST images with a resolution of ${\sim} 0.1\arcsec$ \citep{arrakis2014} a population of smaller central rings can be identified. Nevertheless, our sample exhibits a similar behavior with more small central rings than larger ones (see \autoref{fig:r_ring_distr}), and a two-sample Kolmogorov–Smirnov test between the AINUR and our measured radii yields a $p_{KS}$-value of $27.3\,\% > 5\,\%$, suggesting that the two samples are not significantly different. Indeed, direct comparison of the radii shows that almost all our radius measurements agree within $2 \sigma$ with the ones by AINUR (compare \autoref{table:ring-props} and Table A2 in \citealt{comeron_ainur_2010}). The exception is the central ring in NGC\,1433, where the CO geometry within the ring region is more complex than just a ring plus a central emission component (see \autoref{fig:rings_gallery}). Therefore, different studies might come to different conclusions about the position of the central ring.

The central ring radii relative to the disk size as measured by the 25th magnitude isophotal \textit{B}-band radius, $r_{25}$, show a similar behavior in this study and in AINUR\footnote{AINUR used the relative ring diameter $D_\text{ring}/D_0$ and find that $D_0$ is close to $D_{25}$ \citep{comeron_ainur_2010}, so $r_\text{ring}/r_{25} \approx D_\text{ring}/D_0$ for most of the galaxies.} with most central rings residing within $r_{25}/100 \le r_\text{ring} \le r_{25}/10$. AINUR reports a peak of the distribution at $\log(r_\text{ring}/r_{25}) \approx -1.3$, while we find the peak at $\approx -1.7$ (\autoref{fig:r_ring_distr}).

In summary, our results for the central ring fractions and radii are consistent with the ones found in the literature, especially from the largest literature sample of central rings, AINUR.

\subsection{Central ring properties along the SFMS}\label{sec:ring_props-SFMS}

From \autoref{fig:rings_on_SFMS} it is already evident that central rings occur preferentially in massive galaxies with stellar masses of $\log(M_*/M_\odot) > 10$ consistent with hydrodynamical simulations \citep{verwilghen_2024} and observations (\citealt{comeron_ainur_2010}, using stellar  mass measurements from the $z=0$ Multiwavelength
Galaxy Synthesis project (z0MGS), \citealt{Leroy_2019}; \citealt{fraser-mckelvie_2020_SFR-bar}). The exception is the early-type galaxy NGC\,4476 with a stellar mass of $\log(M_*/M_\odot) = 9.81$ which is still close to $10^{10}\,M_\odot$. Furthermore, being an ETG, it exhibits different dynamics than (barred) spirals and its central ring might thus not be comparable with the other ones\footnote{NGC\,4476's central ring is also an outlier in the molecular gas fraction, where it contains $60\,\%$ of its host galaxy's molecular gas.} (\S \ref{sec:ring-comments}).

Furthermore, there is a correlation between $M_\text{mol}^\text{ring}$ and $M_*^\text{gal}$ with a Pearson correlation coefficient\footnote{Note the Pearson correlation coefficients are calculated for the quantities themselves (not their logarithms) unless stated otherwise. Errors on $p$-values are calculated by resampling the data $1000$ times within their error bars and taking the standard deviation of the resulting $p$-value distribution.} of $\rho_\text{Pearson}=0.86$ with a corresponding $p$-value of $p < 10^{-4}$ which is thus statistically significant: The central rings with more molecular gas are found in more massive galaxies. This is not surprising as more massive galaxies generally tend to harbor more massive structures.

\begin{figure}[t!]
    \centering
    \includegraphics[width=0.48\textwidth]{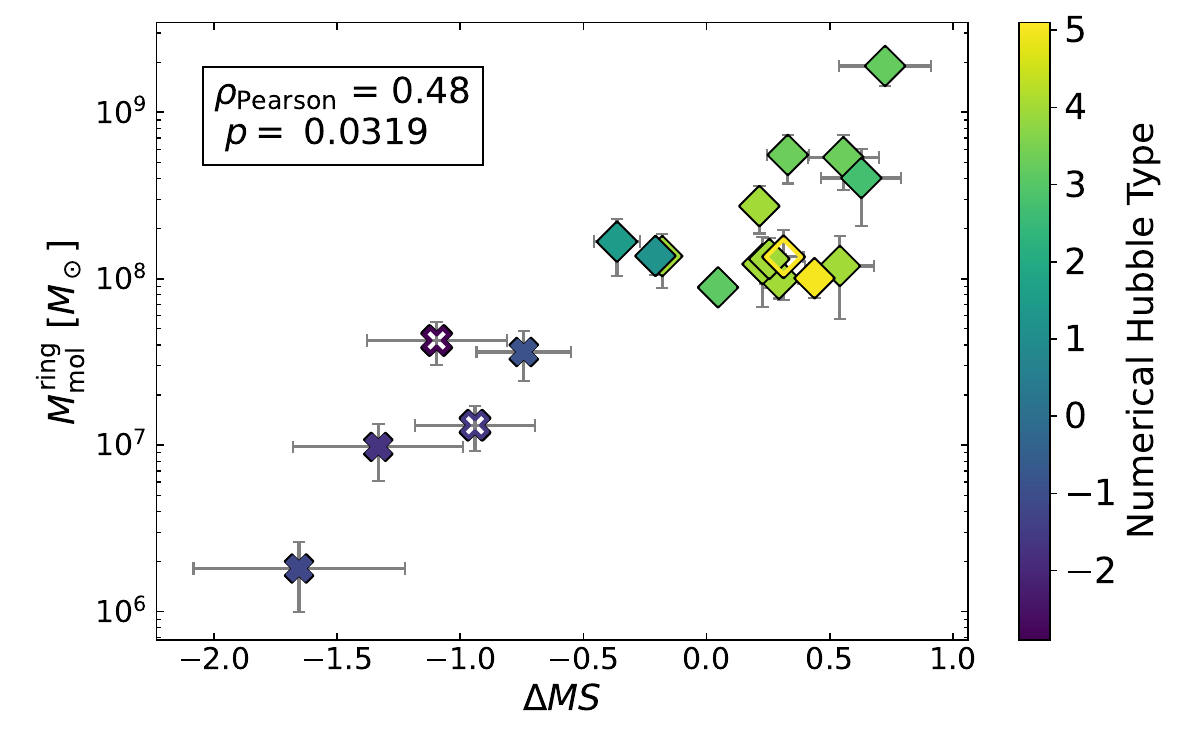}
    \caption{Dependency of the molecular gas mass in PHANGS central rings on the offset from the main sequence $\Delta MS$. The symbols follow \autoref{fig:rings_on_SFMS} (barred spiral = diamond, unbarred spiral = open diamond, barred ETGs = cross, and unbarred ETGs = open cross). The Pearson correlation coefficient is listed in the top left. Data points are color-coded according to the host galaxies' numerical Hubble Type.}
    \label{fig:M_mol-vs-delta_ms}
\end{figure}

Also the offset from the main sequence $\Delta MS$ and $M_\text{mol}^\text{ring}$ are statistically significant correlated with a Pearson-correlation coefficient of $\rho_\text{Pearson} = 0.48$ and the corresponding $p$-value of $p = (3.2 \pm 1.8)\,\%$ (see \autoref{fig:M_mol-vs-delta_ms}): The central rings with little molecular gas are found below the SFMS, while the ones with more molecular gas are found above it. This correlation is likely inherited from a correlation between $M_\text{mol}^\text{gal}$ and $\Delta MS$ (with $\rho_\text{Pearson} = 0.74$ and $p = (0.02\pm 0.02)\,\%$), as we find a relatively flat $M_\text{mol}^\text{ring}/M_\text{mol}^\text{gal}$ across our sample (with few exceptions, see \autoref{fig:delta_ms}).

We find no reliable correlation between $\Delta MS$ and the molecular gas fraction within the ring ($M_\text{mol}^\text{ring} / M_\text{mol}^\text{gal}$) or between $\Delta MS$ and the ring depletion time (\autoref{app:further_biases}).

Here we note that the small number of central rings in our study limits the conclusions that can be drawn. Especially the limited sample of $14$ central rings with SFR maps caused further correlations we considered, e.g. between the ring's molecular gas masses and their SFRs, not to be reliable.

\subsection{Bar parameters and $M_\text{mol}^\text{ring}$}\label{sec:M_mol-influences}

\begin{figure*}
    \centering
    \includegraphics[width=0.9\textwidth]{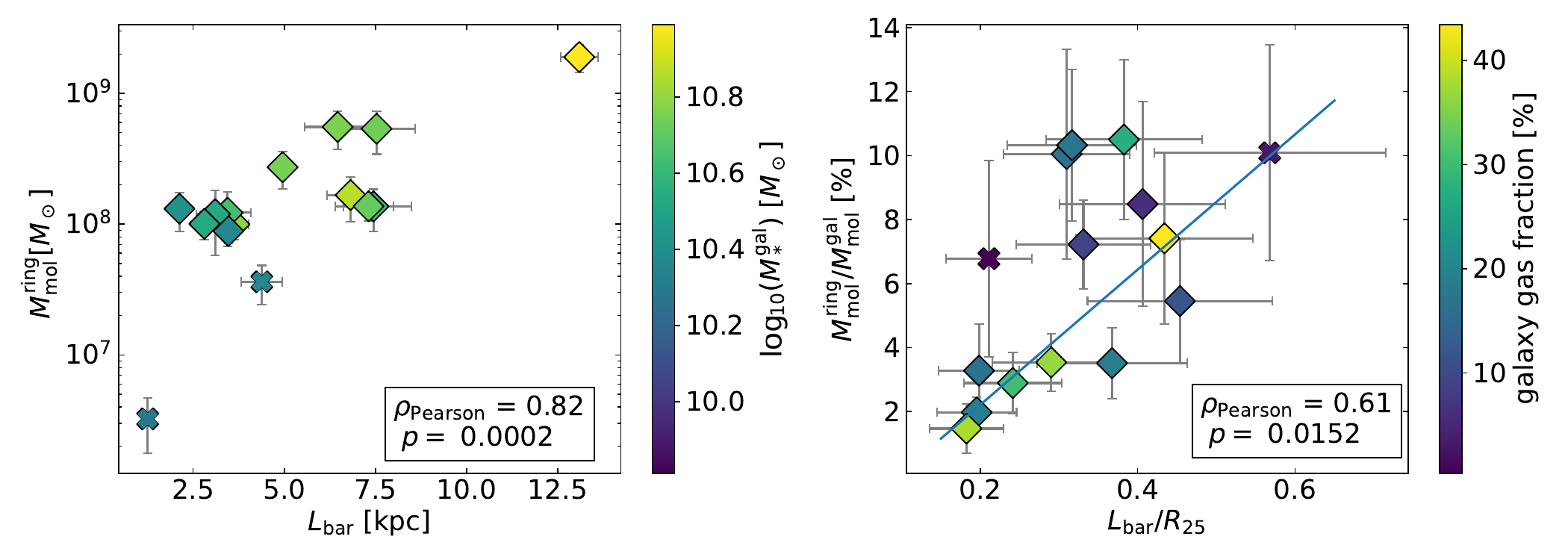}
    \caption{Left: Dependency of the molecular gas mass in PHANGS central rings on the bar length. Data points are color-coded by their host galaxies' stellar mass $M_*^\text{gal}$. Right: Dependency of the relative ring molecular gas content on the relative bar length, color-coded by their host galaxies' gas fraction $(M_\text{mol} + M_\text{atomic}^\text{gal})/M_*^\text{gal}$. The blue line is a linear fit to the data, obtained with the \texttt{python} tool \texttt{scipy.optimize.curve\_fit}. In both panels, the symbols follow \autoref{fig:rings_on_SFMS}  (barred spiral = diamond, barred ETGs = cross) and the Pearson correlation coefficients are provided in the bottom right. }
    \label{fig:frac_mol-vs-L_bar_rel}
\end{figure*}

Theory and simulations (e.g., \citealt{athanassoula_1992, sormani_2015_bar-flow, sormani_2024, verwilghen_2024}) imply that bars funnel gas to the center of galaxies, which has also been corroborated observationally \citep[e.g.,][]{sakamoto_1999_bar-inflow, garcia-burillo_05_nuga, sheth_2005_bar-inflow, querejeta_2016_bar-inflow}. Due to the interplay of (bar) non-axisymmetric potential, orbital angular momentum, and gas pressure forces, the gas settles into the ring-like structures studied here. Therefore, it is conceivable to expect a correlation between bar parameters and the molecular gas content of central rings.

For the $15$ galaxies with measured bar parameters \citep{herrera-endoqui_catalogue_2015, diaz-garcia_characterization_2016}, we find a positive correlation between the bar length (semi-major axis) and the molecular gas content (\autoref{fig:frac_mol-vs-L_bar_rel}, left panel). Thus, we find indeed that central rings with a larger molecular gas mass are found within longer bars. However, this correlation is likely inherited from a correlation between bar length and host galaxy stellar mass \citep{erwin_2019_Lbar} and between galaxy stellar mass and central ring molecular gas mass (\autoref{fig:rings_on_SFMS}, \S \ref{sec:ring_props-SFMS}).
In order to understand the central ring funneling mechanisms independently of global galaxy scaling relations, it is more useful to consider relative parameters: When considering the central ring molecular gas fraction $M_\text{mol}^\text{ring} / M_\text{mol}^\text{gal}$ as a function of the relative bar length to the disk size $L_\text{bar} / r_{25}$, we obtain a Pearson correlation coefficient of $\rho_\text{Pearson} = 0.61$ with a corresponding $p$-value of $p = (1.5 \pm 9)\,\%$, which suggests that this correlation is not very stable against resampling within the error bars. We find no significant influence of stellar mass $M_*^\text{gal}$ of the galaxy or the galaxy gas fraction $(M_\text{mol}^\text{gal} + M_\text{atomic}^\text{gal}) / M_*^\text{gal}$ on this (moderate) correlation. Similarly, no correlations emerge between the relative\footnote{As all bar strength parameters are relative parameters, either giving the relative torque ($Q_\text{b}$), or the contrast between bar and the surrounding ($A_2^\text{max}$, $\varepsilon_\text{bar}$), we test for correlations with the relative gas content.} molecular gas content of the rings $M_\text{mol}^\text{ring} / M_\text{mol}^\text{gal}$ and the classical parameters for bar strength ($p>5\,\%$; see figures in \autoref{app:further_biases}).

\section{Discussion}\label{sec:discuss}
\subsection{Molecular gas as tracer for central rings}\label{secc:disc:mol-tracer}

The similar central ring fraction reported in this paper and in the literature using HST images and H recombination lines \citep{knapen_2005, comeron_ainur_2010} implies that molecular gas traces central rings equally well as star-formation, dust, and ionized gas (all used by \citealt{comeron_ainur_2010}, the latter also by \citealt{knapen_2005}) which is seen in UV imaging, color index and structure maps, or Pa$\alpha$ and H$\alpha$, respectively. This is corroborated by the similar radius distributions of central rings found in this study and AINUR. The comparison between different tracers has become possible due to the unprecedented resolution of CO(2-1) maps taken with ALMA, reaching a similar resolution than only optical observations (e.g. ground-based narrowband H$\alpha$) before.

We conclude that the choice of tracer (molecular lines, H recombination lines, optical and near-IR HST images) does not influence the detection and visual identification of central rings.

Having said this, we speculate that there are other effective ways of detecting molecular central rings in nearby galaxies. Especially JWST allows for using dust or PAH (Polycyclic Aromatic Hydrocarbon) emission to trace galaxy structures at high resolution. Given the smoother appearance of central rings in such maps, the application of automated radial flux analysis and central ring geometry identification can allow for more consistent determination than with the visual detection adopted here. Upcoming work by Neumann (in prep.) and Emsellem (in prep.) will further explore steps in this direction.

\subsection{The MW CMZ is not a special central ring}\label{sec:disc:ring-props}

This study confirms the notion of central rings as extreme environments for star formation \citep[e.g.,][]{schinner_2024}: They have relative radii of $0.01 {-} 0.1\,r_{25}$ (both in this work and in \citealt{comeron_ainur_2010}), which corresponds roughly to an area of $0.01 {-} 1\,\%$ of their host galaxy’s disk, but contain normally ${\sim} 1 {-} 10\,\%$ of the galaxy’s molecular gas reservoir and contribute ${\sim} 1 {-} 25\,\%$ of the star formation rate. This agrees with earlier observations \citep[e.g.,][]{comeron_ainur_2010, querejeta_2021} and simulations \citep[e.g.,][and many others]{athanassoula_1992, seo_2013_ring-hydro-sim, armillotta_2019_CMZ, moon_2021_ringSFR_sim, verwilghen_2024}, that find enhanced molecular gas surface densities and SFRs in the centers of galaxies. Also the average central ring depletion time (mostly $< 1\,$Gyr) is shorter than the typical depletion time in galaxy discs (normally ${\sim} 1 {-} 2\,$Gyr), as also found in previous observational studies \citep[e.g.][]{querejeta_2021, teng_2024_sfe, leroy_2025_sfe}.

We find that when compared to central rings in PHANGS spiral galaxies, the MW CMZ is smaller and less massive, but apart from this it is not special: It exhibits similar star-formation properties (SFR fraction and $t_\text{dep}$) and hints toward a similar funneling dynamics by the bar (molecular gas fraction) than extragalactic central rings.

This is interesting as many studies suggest that the MW CMZ is under-producing stars \citep[e.g.,][and references within]{longmore_2013, kruijssen_2014_CMZ_SFR, Henshaw_2023_CMZ}: While it roughly falls on the Schmidt-Kennicutt relation \citep[depending on its assumed geometry, for details see][]{Henshaw_2023_CMZ}, it produces by an order of magnitude fewer stars than expected from the so-called dense gas star formation relation \citep{gao_solomon_2004, longmore_2013, kruijssen_2014_CMZ_SFR, barnes_2017_CMZ_SFR}. Processes such as turbulence, magnetic fields or tidal shear are thought to suppress the formation of stars from the dense molecular gas present in the MW CMZ \citep[e.g.,][]{kauffmann_2013_virial_SFR, krumholz_2015_ringSFR, mills2017milky, moon_2023_magfields, Henshaw_2023_CMZ}. Given the clear similarities between extragalactic central rings and the MW CMZ, we suggest that such processes are at play in PHANGS central rings as well.

\subsection{The role of bars and $\alpha_\text{CO}$}\label{sec:disc:bars-funneling}

The importance of bars for central ring formation in theory and simulation is underlined by our study, where most central rings reside in barred galaxies: $17$ out of $20$ central rings are found within bars\footnote{See Sect.\,\ref{sec:ring-comments} and \autoref{app:ring_sample}. As the classification of NGC\,2997 as unbarred is not entirely clear, this fraction might even be $18/20 = 90\,\%$.}. Using $15$ galaxies with bar parameters available \citep{herrera-endoqui_catalogue_2015, diaz-garcia_characterization_2016}, we find that longer bars contain more massive rings (\autoref{fig:frac_mol-vs-L_bar_rel}, left panel). There is only tentative evidence for a correlation between relative bar length and fraction of molecular gas within the central rings (\autoref{fig:frac_mol-vs-L_bar_rel}; $p = (1.5 \pm 9)\,\%$) as the relatively large uncertainties for $M_\text{mol}^\text{ring} / M_\text{mol}^\text{gal}$ and $L_\text{bar} / r_{25}$ significantly affect our derived $p$-value uncertainties. Our conservative approach to determine the uncertainty on ring properties (see Sect.\,\ref{sec:ring-shapes}) might have in particular overestimated the uncertainty (assumed to be a $1\sigma$ standard deviation) on $M_\text{mol}^\text{ring}$. Thus, a more accurate determination might confirm a correlation.

Another factor influencing the correlation is the adopted conversion factor: Using the \citet{bolatto_alpha_2013} $\alpha_\text{CO}$ prescription instead of the \citet{schinner_2024} one, changes the Pearson correlation coefficient to $\rho_\text{Pearson} = 0.74$ with a corresponding $p$-value of $p = (0.2 \pm 3.6)\,\% < 5\,\%$. Therefore, the adopted conversion factor has a relevant effect on the measured correlations, although different $\alpha_\text{CO}$ prescriptions agree within their error bars with each other (${\sim}0.2{-}0.3\,$dex, \citealt{sun_2025_alphaCO}).

The moderate correlation between relative bar length and central ring molecular gas fraction hints at a (simplified) central ring fueling scenario: If the bar is longer, it sweeps out a larger area of the galactic disk and gathers more gas within the disk. This gas can be funneled to the center and might lead to the higher central ring molecular gas mass fraction.

However, this notion neglects several dynamical effects that are important for ring fueling: First, the central ring size is thought to be connected with the $x_2$ orbits, which scale both with bar length and the central mass concentration (i.e. the total mass within the center region). When examining this, we find no correlation between ring molecular gas mass and size ($\rho_\text{Pearson}=0.45$, $p = (5.4\pm 6.6)\,\%$) and only a moderate correlation between bar length and ring size ($\rho_\text{Pearson}=0.58$, $p = (2.4\pm 3.3)\,\%$). Thus, we exclude this theoretical scaling between ring size and ring gas mass to cause the $M_\text{mol}^\text{ring} / M_\text{mol}^\text{gal}$ - $L_\text{bar} / r_{25}$ correlation.

Another crucial factor is time evolution, as ring fueling is not constant, but rather subject to episodic accretion \citep[e.g.][for M83 and the MW CMZ]{callanan_2021_m83_cmz, sormani_2019_cmz_inflow}. Furthermore, bars live longer than a bar rotation \citep{sa-freitas_2025_TIMER} and therefore, the amount of gas that can be funneled inwards most likely depends on the gas reservoir at (and beyond) the bar ends -- and not on only the gas mass in the disk region swept up by the bar, as we propose in this simplified ring fueling scenario. We conclude that central ring fueling is more complex, but that bar length likely plays a role.

While there are theoretical expectations that bar strength regulates the molecular gas content in galaxy centers, i.e. stronger bars are believed to funnel gas more efficiently to the center \citep[e.g.,][]{schwarz_bars_1981, sellwood_bars_1993}, we find no correlation between the classical bar strength parameters and the central ring molecular gas mass fraction. Thus, we cannot confirm the notion of stronger bars funneling gas more efficiently to the center.

\subsection{Central rings on the SFMS - Stellar mass limit on ring formation}\label{sec:disc:SFMS-stellar_mass}

Hydrodynamical simulations of PHANGS-like galaxies show a minimum galaxy stellar mass for the formation of central rings, or central gas reservoirs \citep{verwilghen_2024, verwilghen_2025}. This is consistent with our findings that central rings almost only appear within galaxies with $M_* > 10^{10}\,M_\odot$\footnote{With the exception of NGC\,4476, which is an outlier also in the ring molecular gas fraction.}. \citet{verwilghen_2024} explain this behavior with the balance of supernovae (SN) feedback and the gravitational potential: In lower mass galaxies, the gravitational potential is not strong enough to counteract the SN feedback driving a gas outflow from the central regions, which leads to a depletion of gas before a ring can be formed. In higher mass galaxies, the gravitational potential overcomes the SN feedback, and central rings can form. 

Interestingly, the central rings' molecular gas mass correlates positively with the offset from the main sequence. As this trend is driven by ETGs (with $\Delta MS < -0.5$) that behave differently than normally star-forming spiral galaxies \citep{Davis_2013_gasETG, williams_2025_ETG}, it is difficult to determine if the trend is due to the different morphology or due to other galaxy properties. 

The central ring depletion time exhibits a different behavior: When including the ETGs, there is no correlation between $\Delta MS$ and $t_\text{dep}$ ($\rho_\text{Pearson}=-0.13$ and $p=(66\pm15)\,\%$), but without ETGs, the $p$-value decreases to $p = (4\pm7)\,\%$. While this correlation is tentative in our data, further investigations would be interesting, especially in light of the model proposed by \citet{tacchella_2016_SFMS} for high-redshift galaxies: galaxies are confined to a narrow ($\pm0.3\,$dex) star-forming main sequence due to cycles of gas depletion and replenishment that cause galaxies oscillating around the main sequence before eventually being quenched. This leads -- especially for galaxy centers -- to gradients of gas fraction, depletion time and gas density across the main sequence. This model could also explain the large range of values of $M_\text{mol}^\text{ring}$, SFR$_\text{ring}$, and $t_\text{dep}^\text{ring}$, spanning more than one order of magnitude for central rings in spiral galaxies (see \autoref{fig:violin-distr}). Future investigations on this require larger samples spanning a wider parameter range in the SFR - $M_*$ diagram and a careful selection on Hubble type.

\subsection{Are we missing central rings?}\label{sec:caveats:missing}

With our method we will miss central rings that have sizes close to or below our resolution limit of ${\sim}1\,\arcsec \lesssim 100\,$pc as well as rings that cannot be well distinguished from potential nuclear CO emission and hence appear as central gas disks without a notable depression inside them. Further our visual identifications requires rings having a reasonably smooth CO geometry.

AINUR \citep[][]{comeron_ainur_2010} find that about $8\,\%$ of their identified rings have sizes below $100\,$pc which is below our detection limit. Comparison to the lower resolution survey ARRAKIS \citep[with $2\,\arcsec$ resolution compared to $0.1\,\arcsec$ resolution;][]{arrakis2014} shows that $56\,\%$ of these rings are no longer detected. Our non-identification of the central ring in NGC\,4826 is due to this resolution limit: Having a radius of $150\,$pc in AINUR, we only see a central molecular disk with a pronounced southern rim, which might turn out as a full ring with higher-resolution data.

In summary, given our bias against preferentially small ($r\lesssim100\,$pc) central rings, there could be up to another $11$ rings present that appear only as central disks in the PHANGS-ALMA sample studied here. As the distributions of ring properties (Figure \ref{fig:violin-distr}) are biased to larger central rings, we speculate that they might extend to lower values of $r_\text{ring}$, $M_\text{mol}^\text{ring}$, and SFR$_\text{ring}$ for a population of smaller rings. Being a small and not very massive ring compared to the current ring sample, the MW CMZ might turn out to sit well within all those property distributions in this case.

\section{Summary and conclusion}\label{sec:conclusion}

To study properties and possible funneling mechanisms of molecular central rings in nearby galaxies, we investigate $81$ nearby main-sequence galaxies from the PHANGS-ALMA survey\footnote{Which have sufficiently high resolution data available.} and identified $20$ central rings in their molecular gas distribution. We utilize $1\,$\arcsec (${\lesssim} 100\,$pc at a distance of $\lesssim 20\,$Mpc) images of CO(2-1) emission tracing the bulk of the molecular gas distribution \citep{Leroy_2021} in these $20$ rings, as well as MUSE observations tracing SFR surface density $\Sigma_{\rm SFR}$ \citep{belfiore_2023} in 14 of these rings. In order to study sizes and molecular gas content in the rings, we visually determine elliptical annular masks based on the molecular gas distribution of the galaxies. We integrate the ring CO luminosity and convert it into the ring molecular gas mass. Similarly, we integrate the central ring's $\Sigma_\text{SFR}$ to estimate their SFR.\\
Our main results are:
\begin{itemize}
    \item Molecular gas is an excellent tracer for identifying central rings. We find a similar ring fraction of $25\pm 4\,$\% as previous studies using different tracers (UV, optical, NIR). The ring size distribution (both absolute and relative to the disk radius) is similar to that reported by AINUR \citep{comeron_ainur_2010}.
    \item Molecular central rings in nearby galaxies span a radius range ($25$th-$75$th percentile) of ${\sim} 250{-}650\,$pc, they contain typical molecular gas masses of ${\sim} 10^{7.87}{-}10^{8.27}\,M_\odot$, corresponding to typical fractions of ${\sim} 3.5{-}10.1\,$\% of their host galaxies' molecular gas reservoir, and have typical SFRs of ${\sim} 0.05{-}0.36\,M_\odot/$yr, corresponding to fractions of ${\sim} 8{-}23\,$\% of their host galaxies' SFR. Central rings have typical depletion times of ${\sim} 0.44{-}0.76\,$Gyr, which is a factor ${\sim} 2{-}4$ lower than typical depletion times seen in galactic disks. Thus, (most) central rings form stars more efficiently than the surrounding disks.
    \item The MW CMZ appears smaller and less massive than PHANGS central rings in spiral galaxies, which might, however, be due to an observational bias to larger rings in the PHANGS sample. Comparing the MW CMZ's molecular gas and SFR fraction (relative to the total MW values) and its depletion time with PHANGS central rings, it seems not to be a special central ring, thus indicating that star formation proceeds similarly across central rings in nearby galaxies.
    \item Central rings have a strong preference for barred galaxies ($17\,/\,20$ rings reside in clearly barred galaxies). Furthermore, longer bars contain more massive central rings, with a moderate correlation also between the relative bar length and the central ring molecular gas fraction $M_\text{mol}^\text{ring} / M_\text{mol}^\text{gal}$. This correlation hints at the bar length being important for the process of ring fueling, although we cannot resolve this highly dynamic process with the data at hand.
    \item Central rings appear (almost) only in high-mass galaxies ($\log(M_* / M_\odot) > 10$), which is in accordance with earlier observations and simulations (\citealt{comeron_ainur_2010} with mass measurements from \citealt{Leroy_2019}; \citealt{fraser-mckelvie_2020_SFR-bar, verwilghen_2024}). Furthermore, ring molecular gas masses are correlated with the offset from the SFMS. However, this trend needs a more careful investigation, as all galaxies, which host central rings and are substantially below the main sequence, are ETGs, that exhibit a different gas dynamics and star-formation (efficiency).
    \item We cannot confirm the notion of strong bars funneling gas more efficiently to the center, as the typical bar strength parameters ($Q_\text{b}$, $\varepsilon_\text{bar}$ and $A_2^\text{max}$) show no correlation with the ring molecular gas fraction.
\end{itemize}
We conclude that central rings are small structures, which preferentially reside in the centers of barred galaxies and that they resemble the MW CMZ in many aspects. Having high molecular gas and SFR contents, they appear as extreme locations for star formation. In the future, it will be interesting to investigate their star formation processes in more detail, using a multi-wavelength approach with data available as part of the still expanding PHANGS survey.

\begin{acknowledgements}
This work was carried out as part of the PHANGS collaboration.\\
S.K.S is supported by a International Research Fellowship of Japan Society for the Promotion of Science (JSPS).
MQ acknowledges support from the Spanish grant PID2022-138560NB-I00, funded by MCIN/AEI/10.13039/501100011033/FEDER, EU.\\
RSK acknowledges financial support from the ERC via Synergy Grant ``ECOGAL'' (project ID 855130),  from the German Excellence Strategy via the Heidelberg Cluster ``STRUCTURES'' (EXC 2181 - 390900948), and from the German Ministry for Economic Affairs and Climate Action in project ``MAINN'' (funding ID 50OO2206).  RSK also thanks the 2024/25 Class of Radcliffe Fellows for highly interesting and stimulating discussions.\\
HAP acknowledges support from the National Science and Technology Council of Taiwan under grant 113-2112-M-032 -014 -MY3. \\
MCS acknowledges financial support from the European Research Council under the ERC Starting Grant ``GalFlow'' (grant 101116226) and from Fondazione Cariplo under the grant ERC attrattivit\`{a} n. 2023-3014.\\
This paper makes use of the following ALMA data: \linebreak
ADS/JAO.ALMA\#2013.1.01161.S, \linebreak 
ADS/JAO.ALMA\#2015.1.00121.S, \linebreak 
ADS/JAO.ALMA\#2015.1.00925.S, \linebreak 
ADS/JAO.ALMA\#2015.1.00956.S, \linebreak 
ADS/JAO.ALMA\#2016.1.00386.S, \linebreak 
ADS/JAO.ALMA\#2017.1.00392.S, \linebreak
ADS/JAO.ALMA\#2017.1.00886.L, \linebreak 
ADS/JAO.ALMA\#2018.1.01651.S. \linebreak 
ALMA is a partnership of ESO (representing its member states), NSF (USA) and NINS (Japan), together with NRC (Canada), MOST and ASIAA (Taiwan), and KASI (Republic of Korea), in cooperation with the Republic of Chile. The Joint ALMA Observatory is operated by ESO, AUI/NRAO and NAOJ.\\
Based on observations collected at the European Southern Observatory under ESO programmes 094.C-0623 (PI: Kreckel), 095.C-0473,  098.C-0484 (PI: Blanc), 1100.B-0651 (PHANGS-MUSE; PI: Schinnerer), as well as 094.B-0321 (MAGNUM; PI: Marconi), 099.B-0242, 0100.B-0116, 098.B-0551 (MAD; PI: Carollo) and 097.B-0640 (TIMER; PI: Gadotti).
\end{acknowledgements}

\bibliography{Bibliography}{}
\bibliographystyle{aa}{}

\newpage

\begin{appendix}
\onecolumn

\section{Central ring sample -- comments on individual rings}\label{app:ring_sample}
In the following we comment on the morphology of the central rings and their host galaxies identified in this work. We adopt the morphology of the host galaxies from the ``Third Reference Catalogue of Bright Galaxies'' \citep[RC3,][]{RC3_1991}. We list the original reference for each central ring, where it has been first reported, following \citet{comeron_ainur_2010} for those rings included both in their work and this study.

\begin{itemize}
\item[]\textbf{NGC\,1097} Barred spiral galaxy, central ring first detected by \citet{burbidge_1962} in H$\alpha$. \citet{buta_1993}, however, claim that the ring is a pseudo-ring. Following AINUR \citep{comeron_ainur_2010}, we regard the ring as a typical central ring: It appears as an almost perfect closed ring in the molecular gas distribution, distinct from an emission component in the very center.

\item[]\textbf{NGC\,1300} Barred spiral galaxy, central ring first mentioned by \citet{pogge_1989}, who detected it in H$\alpha$. Closed, but asymmetric ring, its northern part is brighter than the southern one.

\item[]\textbf{NGC\,1365} Barred spiral galaxy, central ring reported by, e.g., \citet{sandqvist_ngc1365_1995}, observing it in the radio continuum. Furthermore, \citet{Schinnerer_2023} studied the molecular gas dynamics of this specific ring in detail (resolution of ${\sim} 0.3\arcsec \simeq 30\,$pc). Rather elliptical (both in projection and deprojection) ring that is brighter in CO along the northeastern rim.

\item[]\textbf{NGC\,1433} Barred spiral galaxy, central ring first reported as ring of concentrated H\,{\sc ii} regions by \citet{buta_1983}. It shows a lot of emission within the central ring, which makes it difficult to determine a clear ring shape. Therefore, AINUR \citep{comeron_ainur_2010} reports a ring of roughly half the size found in this study. CO emission along the ring appears irregular and clumpy, but the ring is almost completely closed.

\item[]\textbf{NGC\,1512} Barred spiral galaxy, central ring first detected by \citet{buta_1988} in H$\alpha$. The CO ring is very bright at the connection points with the bar lanes and fainter in between, but it is completely closed and very distinct from a nuclear emission component.

\item[]\textbf{NGC\,1566} Weakly barred spiral galaxy, central ring first directly detected ``as a dusty feature in the color-index map'' by \citet{comeron_ainur_2010}. The molecular ring is made up of a few diffuse features that do not connect to form a full ring. Dusty features seen by JWST at $7.7\,\mu$m \citep{Lee_jwst_2023}, however, clearly show a full ring.

\item[]\textbf{NGC\,1672} Barred spiral galaxy, central ring first detected as ring of H\,{\sc ii} regions by \citet{diaz_1999}. Closed but asymmetric ring, the south is brighter than the north in CO emission.

\item[]\textbf{NGC\,2566} Barred spiral galaxy, central ring reported by \citet{carollo_2002} as nuclear star-forming spiral\footnote{\citet{carollo_2002} do not provide a size of this nuclear star-forming spiral which we could directly compare with our central ring measurement. As we do not see any spiral feature in CO(2-1) emission, we think the features are actually the same, but may look different in the respective tracers.} in optical-NIR color maps. In CO, the ring is brighter in the south and almost blurs with a central emission component.

\item[]\textbf{NGC\,2903} Barred spiral galaxy, central ring first detected by \citet{sakamoto_1999} in CO($J = 1\rightarrow0$) emission. Diffuse molecular features around the central ring complicate defining the exact shape of the ring. Almost closed central ring, which appears very elliptical due to the high inclination of the host galaxy.

\item[]\textbf{NGC\,2997} Weakly barred spiral galaxy, central ring first detected by \citet{sersic_1965} in the optical. The classification of this galaxy as either barred or unbarred is debated: While \citet{comeron_ainur_2010} regards NGC\,2997 as unbarred, \citet{Elmegreen_1999} adopt the classification in the RC3 of this galaxy as weakly barred. Instead of a bar, \citet{comeron_ainur_2010} suggest that the central ring might be induced by NGC\,2997's strong spiral arms. In the plots throughout this paper, we mark NGC\,2997 as unbarred spiral. The CO emission ring has a gap in the west close to the connection point with the western spiral arm.

\item[]\textbf{NGC\,3351} Barred spiral galaxy, central ring first reported by \citet{sersic_1965} in the optical. Almost completely closed ring in CO emission, which is distinct from an emission component in the very center.

\item[]\textbf{NGC\,3489} Weakly barred ETG, central ring reported by, e.g., \citet{erwin_2002}, who observed it in the NIR. Given the galaxy's high inclination it was not considered by AINUR \citep{comeron_ainur_2010}. Very elliptical ring in CO emission due to the inclination. It is roughly broken into a western and eastern half.

\item[]\textbf{NGC\,4303} Weakly barred spiral galaxy, central ring first detected by \citet{moellenhoff_2001} in the NIR. The brighter western part of the molecular central ring blurs with emission associated with the nucleus. Therefore, the molecular ring seems to be offset from the star formation ring which is centered at the nucleus.

\item[]\textbf{NGC\,4321} Weakly barred spiral galaxy, central ring first detected by \citet{sersic_1965} in the optical. According to \citet{comeron_ainur_2010}, this ring is a limiting case between a classical central ring and a pseudo-ring. Indeed, in CO(2-1) emission, the ring is broken into a western and eastern half, that are closely connected to the bar lanes, which we still consider as ring-like. The ring is clearly distinct from a nuclear emission component.

\item[]\textbf{NGC\,4459} ETG, containing two central rings \citep{comeron_ainur_2010}. The feature identified as central ring in this paper was first detected as ``inner nuclear ring''  in HST UV images by \citet{comeron_ainur_2010}. They suggest that the central ring is formed by interactions with the neighboring galaxies NGC\,4468 and/or NGC\,4474. Very smooth and closed thick ring in CO, no central emission component.

\item[]\textbf{NGC\,4476} ETG, central ring reported by \citet{prugniel_ngc4476_1987} as dust ring in the optical and NUV. The central ring is the only feature seen in the molecular gas distribution of this galaxy. Completely closed ring, no nuclear emission component.

\item[]\textbf{NGC\,4477} Barred ETG, central ring reported by \citet{crocker_ngc4477_2011}, who observed it in CO($J = 1\rightarrow0$) and CO($J = 2\rightarrow1$) emission. The completely closed central ring and the beginnings of the bar lanes are the only features seen in NGC\,4477's molecular gas distribution.

\item[]\textbf{NGC\,5236} Weakly barred spiral galaxy, central ring first detected by \citet{buta_1993} in the optical. Almost completely closed ring in CO with bright connection points to the bar lanes. Some diffuse features around the ring.

\item[]\textbf{NGC\,5248} Weakly barred spiral galaxy, containing two central rings. The feature identified as central ring in this paper is the ``outer nuclear ring'' \citep{comeron_ainur_2010}, which was first detected by \citet{sersic_1965} in the optical. The bar lanes near the connection points with the central ring are very bright in CO(2-1) emission and might be confused as being part of the ring. Compared to the bar lanes and the nucleus, the ring is rather faint in CO and a bit clumpy.

\item[]\textbf{NGC\,7743} Barred ETG, central ring reported by \citet{martini_ngc7743_2003} as loosely wound nuclear spiral in the optical and NIR. The central ring and few diffuse features around it are the only features seen in NGC\,7743's molecular gas distribution. The ring is almost completely closed, but has faint spots in the north and south.

\end{itemize}

\section{Central ring masks}\label{app:ring-masks}

{
\begin{longtable}{cccccccccccccc}
\caption{Central ring masks. \label{table:app:ring-masks}}\\ 
\noalign{\smallskip}
\hline
\hline
\noalign{\smallskip}
Galaxy & $h_\text{in}^\text{best}$ & $w_\text{in}^\text{best}$ & $h_\text{out}^\text{best}$ & $w_\text{out}^\text{best}$ & $h_\text{in}^\text{broad}$ & $w_\text{in}^\text{broad}$ & $h_\text{out}^\text{broad}$ & $w_\text{out}^\text{broad}$ & $h_\text{in}^\text{strict}$ & $w_\text{in}^\text{strict}$ & $h_\text{out}^\text{strict}$ & $w_\text{out}^\text{strict}$ & PA$^\text{ring}$ \\
& [\arcsec] & [\arcsec] & [\arcsec] & [\arcsec] & [\arcsec] & [\arcsec] & [\arcsec] & [\arcsec] & [\arcsec] & [\arcsec] & [\arcsec] & [\arcsec] & [$\degree$] \\
(1) & (2) & (3) & (4) & (5) & (6) & (7) & (8) & (9) & (10) & (11) & (12) & (13) & (14)\\ 
\hline
\endfirsthead
NGC\,1097 & 12.1 & 10.1 & 26.9 & 24.8 & 10.1 & 8.63 & 29.3 & 27.7 & 14.6 & 12.8 & 24.1 & 22.1 & 30.0 \\
NGC\,1300 & 4.18 & 2.98 & 12.3 & 9.96 & 3.74 & 2.49 & 13.3 & 11.2 & 4.75 & 3.57 & 10.6 & 8.82 & 130.0 \\
NGC\,1365 & 8.56 & 4.33 & 26.1 & 14.2 & 5.93 & 3.03 & 28.6 & 17.5 & 12.5 & 5.79 & 22.6 & 13.1 & 25.0 \\
NGC\,1433 & 18.9 & 11.8 & 28.7 & 22.8 & 9.45 & 7.80 & 30.7 & 23.1 & 19.7 & 12.6 & 28.0 & 20.9 & 20.0 \\
NGC\,1512 & 8.81 & 8.10 & 20.3 & 16.4 & 6.88 & 5.09 & 22.9 & 18.2 & 12.1 & 8.98 & 18.4 & 14.8 & 105.0 \\
NGC\,1566 & 11.6 & 10.1 & 26.9 & 20.6 & 10.2 & 9.62 & 28.3 & 21.4 & 14.3 & 11.8 & 24.4 & 19.8 & 15.0 \\
NGC\,1672 & 6.99 & 6.18 & 19.5 & 17.3 & 5.99 & 5.08 & 21.9 & 20.4 & 8.22 & 7.24 & 17.9 & 16.2 & 125.0 \\
NGC\,2566 & 2.88 & 2.63 & 7.74 & 7.65 & 2.48 & 1.98 & 8.54 & 8.32 & 3.30 & 2.76 & 7.13 & 6.61 & 130.0 \\
NGC\,2903 & 4.03 & 1.73 & 15.8 & 9.32 & 3.34 & 1.03 & 20.1 & 10.3 & 5.49 & 2.05 & 14.3 & 7.09 & 14.0 \\
NGC\,2997 & 5.27 & 4.54 & 15.3 & 13.9 & 3.85 & 3.35 & 17.0 & 14.6 & 5.98 & 5.72 & 13.4 & 12.7 & 45.0 \\
NGC\,3351 & 9.59 & 4.72 & 20.4 & 12.4 & 6.03 & 3.88 & 21.9 & 13.1 & 10.2 & 5.66 & 18.0 & 10.6 & 23.0 \\
NGC\,3489 & 3.38 & 1.05 & 7.03 & 3.37 & 2.56 & 0.702 & 6.77 & 4.09 & 3.96 & 1.18 & 6.10 & 3.04 & 65.0 \\
NGC\,4303 & 2.95 & 2.12 & 10.95 & 9.06 & 3.44 & 0.678 & 13.3 & 10.8 & 5.19 & 2.11 & 10.5 & 7.40 & 165.0 \\
NGC\,4321 & 8.63 & 7.81 & 20.5 & 19.1 & 6.70 & 5.89 & 26.6 & 20.4 & 12.5 & 9.74 & 18.9 & 16.8 & 0.0 \\
NGC\,4459 & 0.744 & 0.640 & 5.15 & 3.68 & 0.590 & 0.428 & 5.51 & 4.19 & 0.894 & 0.806 & 4.84 & 3.25 & 85.0 \\
NGC\,4476 & 5.75 & 2.38 & 15.7 & 8.87 & 3.14 & 1.40 & 18.3 & 10.2 & 7.52 & 2.76 & 14.9 & 7.18 & 30.0 \\
NGC\,4477 & 1.19 & 0.622 & 4.23 & 2.87 & 0.894 & 0.372 & 5.09 & 3.35 & 1.35 & 0.772 & 3.70 & 2.73 & 95.0 \\
NGC\,5236 & 11.2 & 8.55 & 24.8 & 22.0 & 10.1 & 8.34 & 28.2 & 23.4 & 13.1 & 9.69 & 23.6 & 19.3 & 170.0 \\
NGC\,5248 & 7.70 & 5.35 & 18.5 & 12.3 & 5.78 & 4.41 & 19.2 & 13.2 & 8.56 & 6.07 & 17.3 & 11.6 & 142.0 \\
NGC\,7743 & 0.850 & 0.800 & 4.39 & 3.53 & 0.482 & 0.412 & 5.10 & 4.18 & 1.15 & 0.898 & 3.94 & 3.11 & 105.0 \\
\hline
    \noalign{\smallskip}
\end{longtable}
}
\tablefoot{{The inner height, inner width, outer height, outer width in arcseconds of the ``best'' masks as explained in Sect.\,\ref{sec:ring-shapes}  are contained in Columns (2) - (5); the ``broad'' masks in Col. (6) - (9); the ``strict'' masks in (10) - (13); and the position angle in degree of all ellipses in Col. (14).\\}}

\section{Possible biases and further ring properties}\label{app:further_biases}

In the following, we provide a set of tests for different biases and potential relations.

\begin{figure}
    \centering
    \includegraphics[width=0.4\textwidth]{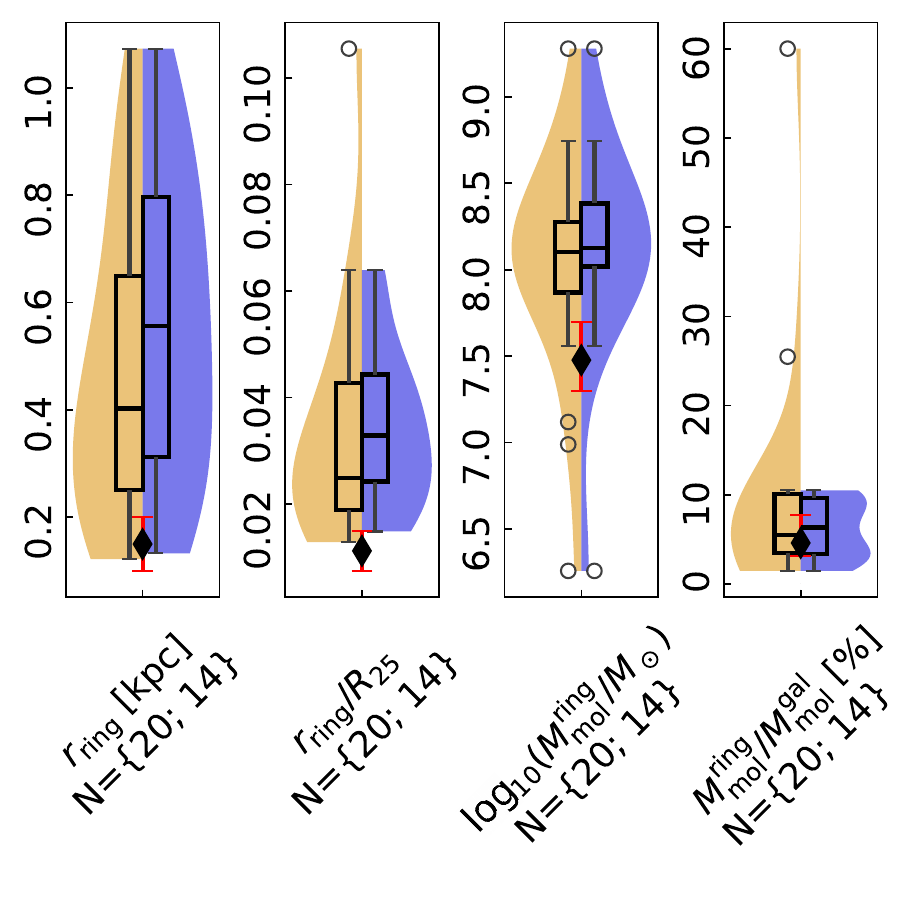}
    \caption{The same plot as \autoref{fig:violin-distr}, but only for $r$, $r/r_{25}$, $M_\text{mol}^\text{ring}$, and $M_\text{mol}^\text{ring}/M_\text{mol}^\text{gal}$, and instead of showing the whole sample and the spirals, the whole sample (left violins) and those with available SFR maps (right violins) are presented. }
    \label{fig:violins_sfr_biased}
\end{figure}

\smallskip
\noindent
\textit{Subsample of rings with SFR maps}:
We verified that our subsample of central rings with SFR maps is representative of the whole sample.
As shown in \autoref{fig:violins_sfr_biased}, these central rings have similar distributions in (relative) radius and molecular gas mass while being slightly larger than the whole sample ($550\pm240\,$pc instead of $400^{+250}_{-150}\,$pc).

\begin{figure}[h]
  \makebox[\textwidth][c]{
    \includegraphics[width=0.49\textwidth]{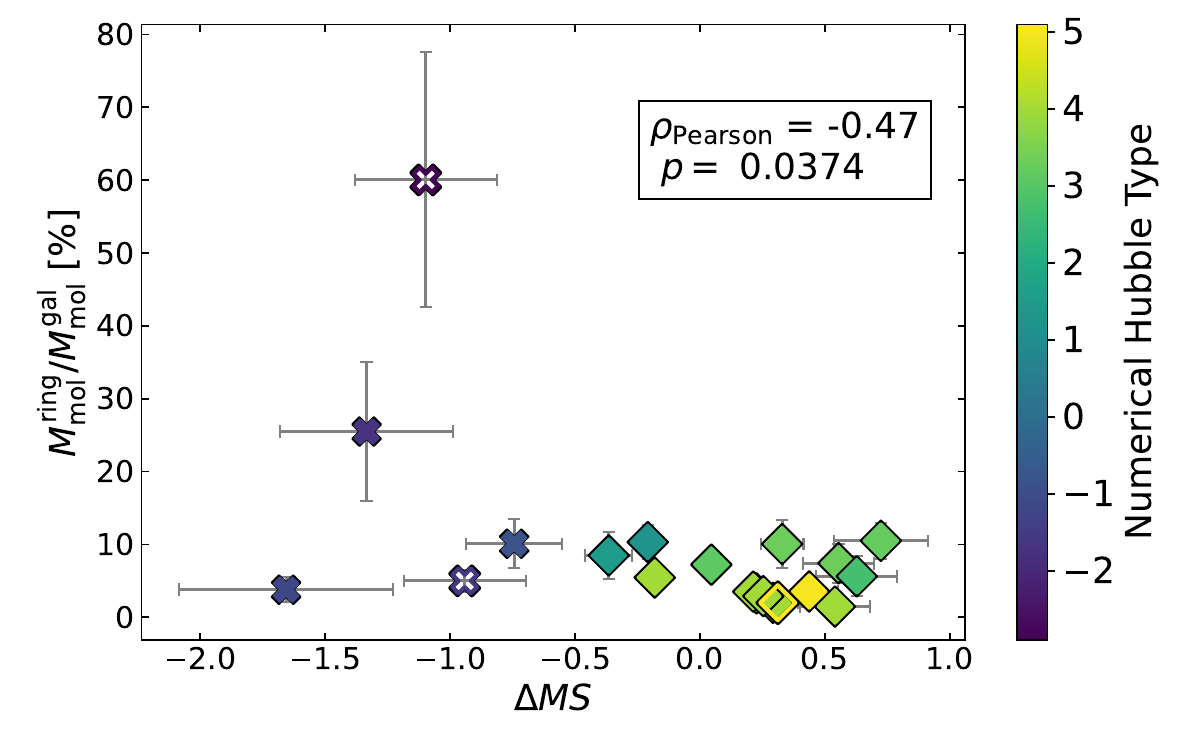}
    \includegraphics[width=0.49\textwidth]{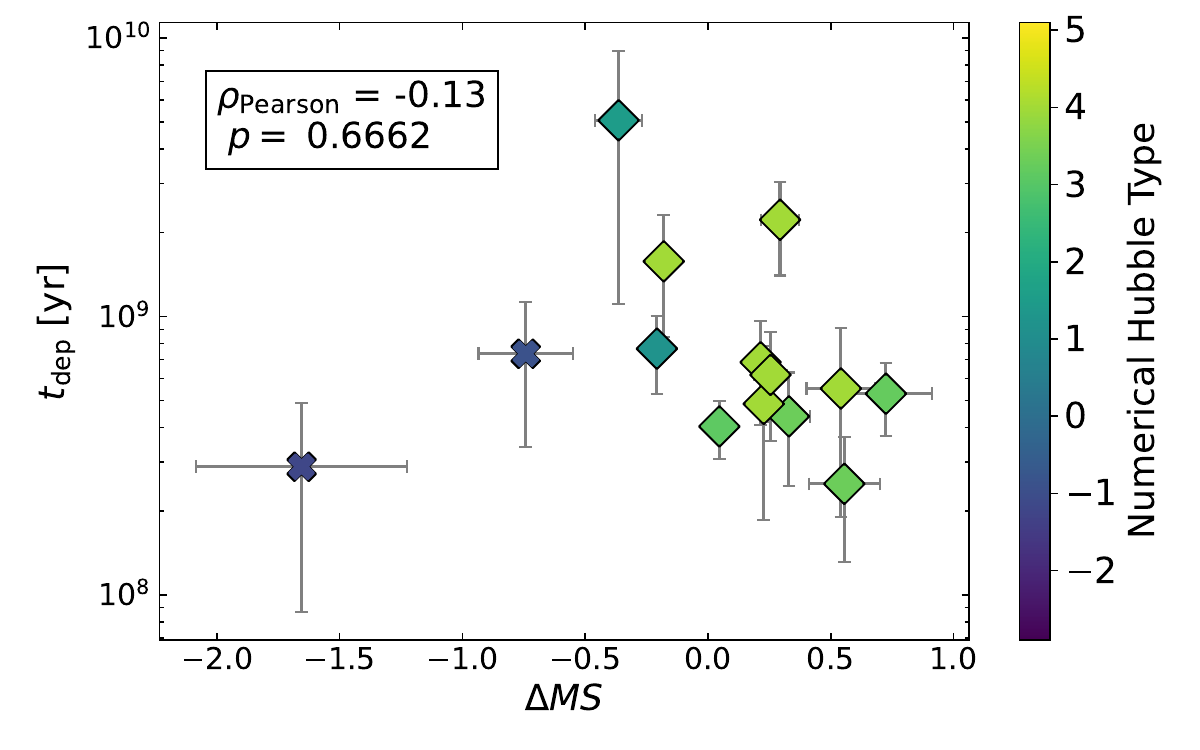}
}
  \caption{Central ring molecular gas fraction (left panel) and their depletion times (right panel) as function of their host galaxies' offset from the main sequence. The symbols are the same as in \autoref{fig:rings_on_SFMS}.}
  \label{fig:delta_ms}
\end{figure}

\smallskip
\noindent
\textit{Correlations with offset from the main sequence}:
In \autoref{fig:delta_ms}, we explore possible correlations between central ring properties and the location of a galaxy with respect to the main sequence. The correlations between the galaxies' offset from the main sequence $\rm \Delta MS$ and the central ring molecular gas mass fraction or their depletion time are ambiguous. The tentative ($p = (3.7 \pm 3.2) \%$) correlation between $M_\text{mol}^\text{ring} / M_\text{mol}^\text{gal}$ and $\rm \Delta MS$ is mostly driven by the outlier NGC\,4476 and disappears if we neglect this source (then, $p = (7 \pm 9) \%$).

\begin{figure}
    \centering
    \includegraphics[width=\linewidth]{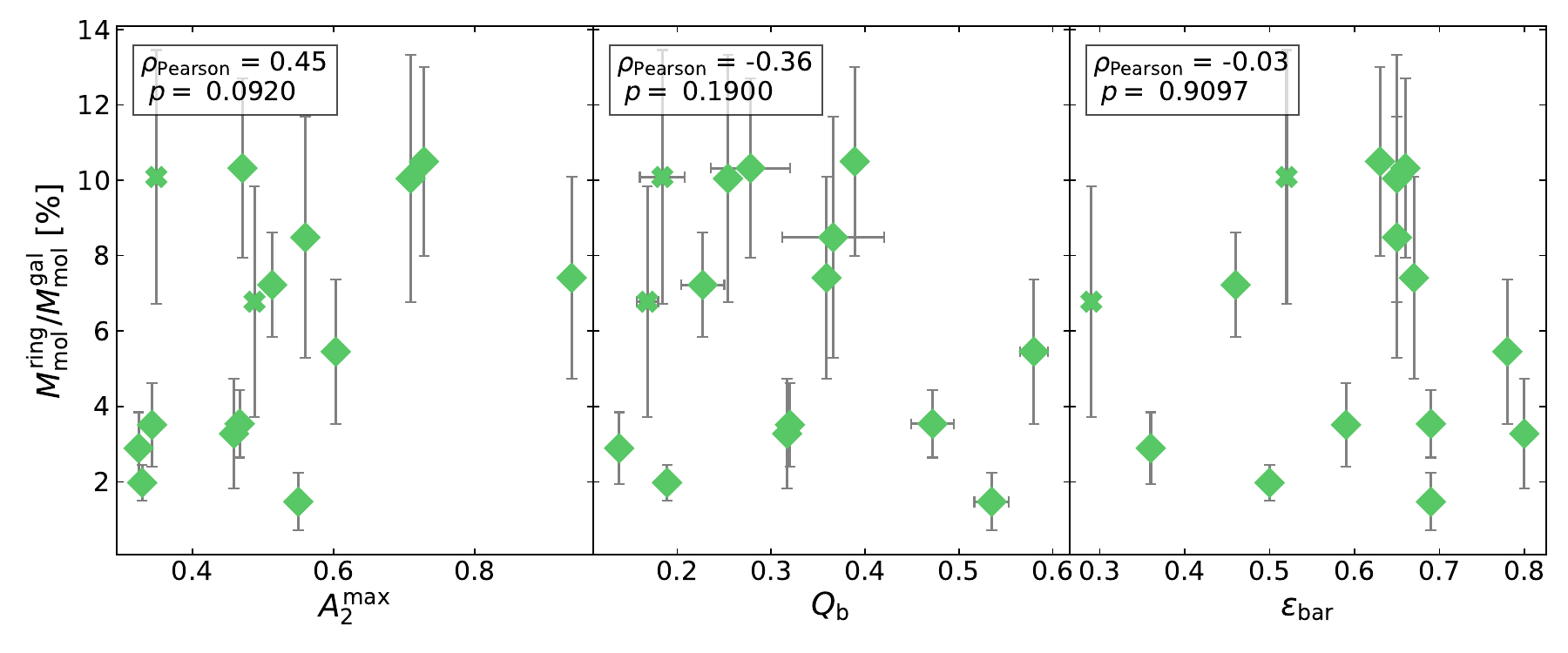}
    \caption{Central ring molecular gas fraction as function of the bar parameters $A_2^\text{max}$ (bar maximum normalized $m=2$ Fourier amplitude; left), $Q_\text{b}$ (bar gravitational torque parameter; middle), and $\varepsilon_\text{bar}$ (bar ellipticity; right). The symbols are the same as in \autoref{fig:rings_on_SFMS}.}
    \label{fig:ring_mol_frac-vs-bar_params}
\end{figure}

\smallskip
\noindent
\textit{Bar strength parameters}:
In \autoref{fig:ring_mol_frac-vs-bar_params}, we include the plots showing the classical bar strength parameters $A_2^\text{max}$, $Q_\text{b}$ \citep{diaz-garcia_characterization_2016} and $\varepsilon_\text{bar}$ \citep{herrera-endoqui_catalogue_2015} as functions of the central ring molecular gas fraction $M_\text{mol}^\text{ring} / M_\text{mol}^\text{gal}$. We do not find any correlation here.

\end{appendix}

\end{document}